\documentclass{JHEP3}
\usepackage{amsmath}
\usepackage{amsfonts}
\usepackage{epsfig}

\def\ap{\alpha^{\prime}}
\def\beq{\begin{equation}}
\def\eeq{\end{equation}}
\def\bra{\langle}
\def\ket{\rangle}
\def\d{\mathrm{d}}
\def\tr{\mathrm{tr}}
\def\ov{\overline}
\def\sst{\scriptscriptstyle}

\newcommand{\fssd}[1]{#1\!\!\!\!/}
\newcommand{\imi}{{\rm i}}
\newcommand{\dbar}{$\overline{\mbox{D}}$}
\newcommand{\dthree}{{\mbox{D3}}}
\newcommand{\dthreebar}{{\overline{\mbox{D}}}{\mbox{3}}}

\author{
S. A. Abel$^a$,
M. D. Goodsell$^b$,
J. Jaeckel$^{a,c}$,
V. V. Khoze$^a$,
A. Ringwald$^d$\\
$^a$Institute for Particle Physics Phenomenology, Durham University, Durham, DH1 3LE, United Kingdom
\\
$^b$Laboratoire de Physique Th\'eorique et Hautes Energies, Tour 24-25, 5eme étage, Boite 126,
4 place Jussieu, F-75252 Paris Cedex 05, France
\\
$^c$Institut f\"ur Theoretische Physik, Universit\"at Heidelberg, Philosophenweg 16, D-69120 Heidelberg, Germany
\\
$^d$Deutsches Elektronen-Synchrotron DESY, Notkestra\ss e 85, D-22607 Hamburg, Germany
\\
E-mail: \email{s.a.abel@durham.ac.uk},
\email{goodsell@lpthe.jussieu.fr},
\email{joerg.jaeckel@durham.ac.uk},
\email{valya.khoze@durham.ac.uk},
\email{andreas.ringwald@desy.de}
}

\title{Kinetic Mixing of
the Photon with Hidden {\boldmath $U(1)$}s in String Phenomenology}

\preprint{
IPPP/08/14\\
DESY 08-026\\
}

\enlargethispage{1.5cm}
\vspace{-1.0cm}
\abstract{
Embeddings of the standard model in type II string theory typically
contain a variety of $U(1)$ gauge factors arising from D-branes in
the bulk. In general, there is no reason why only one of these - the
one corresponding to weak hypercharge - should be massless.
Observations require that standard model particles must be neutral
(or have an extremely small charge) under additional massless
$U(1)$s, i.e. the latter have to belong to a so called hidden
sector. The exchange of heavy messengers, however, can lead to a
kinetic mixing between the hypercharge and the hidden-sector
$U(1)$s, that is testable with near future experiments. This provides
a powerful probe of the hidden sectors and, as a consequence, of
the string theory realisation itself. In the present paper, we show,
using a variety of methods, how the kinetic mixing can be derived
from the underlying type II string compactification, involving
supersymmetric and nonsupersymmetric configurations of D-branes,
both in large volumes and in warped backgrounds with fluxes. We
first demonstrate by explicit example that kinetic mixing occurs in
a completely supersymmetric set-up where we can use conformal field
theory techniques. We then develop a supergravity approach which
allows us to examine the phenomenon in more general backgrounds,
where we find that kinetic mixing is natural in the
context of flux compactifications. We discuss the phenomenological
consequences for experiments at the low-energy frontier, searching
for signatures of light, sub-electronvolt or even massless
hidden-sector $U(1)$ gauge bosons and minicharged particles.}

\begin{document}

\section{Introduction}

Many extensions of the standard model (SM) contain hidden sectors that have no
renormalizable interactions with SM particles. Notably, realistic embeddings of
the standard model in $E_8\times E_8$ heterotic closed string theory as well as in
type I, IIA, or IIB open string theory with branes, often require the existence of hidden
sectors for consistency and for supersymmetry breaking\footnote{For reviews which emphasize the
occurence of hidden sectors in the context of string phenomenology,
see e.g. Refs.~\cite{Quevedo:2002fc,Abel:2004rp,Lust:2004ks,Marchesano:2007de}.}.

At the quantum level, hidden-sector particles will interact with SM particles through the
exchange of massive messengers that couple to both the hidden and visible sectors,
and this can lead to detectable traces of hidden sector physics.
A unique window to hidden sectors is provided by hidden Abelian gauge bosons.
In fact, hidden sector gauge groups often contain $U(1)$ gauge factors which generically
mix kinetically~\cite{Holdom:1985ag,Foot:1991kb}
with the hypercharge $U(1)$ of the visible sector, leading
to terms in the low-energy effective Lagrangian of the form
\beq
\mathcal{L} \supset -\frac{1}{4g_a^2} F^{(a)}_{\mu \nu} F^{\mu \nu}_{(a)}
- \frac{1}{4g_b^2} F^{(b)}_{\mu \nu} F^{\mu \nu}_{(b)}
+ \frac{\chi_{ab}}{2 g_a g_b} F_{\mu \nu}^{(a)} F^{(b) \mu \nu} + m_{ab}^2 A^{(a)}_{\mu} A^{(b)\mu},
\label{LagKM}
\eeq
where
$a(b)$ labels the visible (hidden) $U(1)$, with
field strength $F^{(a(b))}_{\mu\nu}$ and gauge coupling $g_{a(b)}$.
The dimensionless kinetic mixing parameter $\chi_{ab}$, appearing in front of
the effective renormalizable operator in Eq.~(\ref{LagKM}), can be generated at an arbitrarily high
energy scale and does not suffer from any kind of mass suppression from the messengers that
induce it. This makes it an extremely powerful probe of high scale physics; its
measurement could provide clues to physics at energies that may never be accessible to colliders.

The mass mixing term $m_{ab}^2$ in Eq.~(\ref{LagKM}) is, in the context of
string theory, usually associated with the St\"uckelberg
mechanism of mass generation for anomalous $U(1)$s
(see, e.g., Refs.~\cite{Antoniadis:2002cs,Anastasopoulos:2003aj,Kiritsis:2003mc,Anastasopoulos:2004ga,Anastasopoulos:2006cz}).
The $m_{ab}^2$ effects were examined recently in the framework of the ``St\"uckelberg $Z^\prime$ model''
in \cite{Kors:2004dx,Cheung:2007ut,Feldman:2007wj}
whereby a massive (typically ${\cal O}({\rm TeV})$) boson (which may
also kinetically mix with the hypercharge) couples to the
standard model particles directly via such a mass mixing, allowing
it to be produced at the LHC;
the large mass accounts for its current invisibility
(see also Refs.~\cite{Holdom:1990xp,Babu:1997st,Chang:2006fp,Kumar:2006gm}).
It is certainly
a very plausible string-inspired model (possibly even a prediction).

Here, following our earlier work~\cite{Abel:2003ue,Abel:2006qt}, we will address the effect and the
generation of the kinetic mixing term $\chi_{ab}$.
We shall propose searching for truly hidden gauge fields
which are anomaly-free and massless.
In the presence of light or massless hidden fermions, this may be detected
thanks to the kinetic mixing generated at loop level.
This is a complementary string-motivated
scenario, potentially providing different information about the
compact space of string theory which may be impossible to ever obtain
directly. An exhaustive study of the predicted size of kinetic mixing
in realistic compactifications of heterotic string theory has been
performed in Ref.~\cite{Dienes:1996zr}.  Type II models were considered
in previous work~\cite{Abel:2003ue,Abel:2006qt} where we examined the mixing in
non-supersymmetric string set-ups between branes and antibranes in
large toroidal volumes and suggested that the non-observation of
kinetic mixing may be able to place bounds on the string scale
in more general scenarios, or alternatively may place a lower bound on the
kinetic mixing to be observed based on the currently favoured string scale.
However a systematic and rigorous study in
the context of type II string models
is still lacking, and this is the goal the present paper pursues.

Why would one expect kinetic mixing to be of interest in the context of type II models?
Kinetic mixing appears in a Lagrangian when massive modes coupling to different $U(1)$s
are integrated out~\cite{Holdom:1985ag,Holdom:1986eq}.
In the type II context,
hidden $U(1)$s arise as D-branes in the bulk that have no intersection with the
branes responsible for the visible sector.
The heavy modes that are integrated out correspond to open strings
stretched between the visible and hidden stacks of branes. This can also be understood in the closed string
channel as mediation by light or massless closed string (i.e. bulk) modes.
The motivation for a comprehensive study in type II theories therefore
derives from the following general observations: in type II string
compactifications, hidden $U(1)$s are ubiquitous, and there is
no reason to expect all of them to be anomalous and hence heavy.
Furthermore, the Ramond sector on intersecting D-branes always yields the massless charged
matter fermions that could make the kinetic mixing detectable.

The type II models can be further subdivided into
two classes depending on how curved the compact space is supposed to be.
First there are models in which the compact space plays the role of a large
quasi-flat bulk volume. These include the D-branes at singularities (so called bottom-up)
models~\cite{Aldazabal:2000sa}.
For the sake of simplicity we will also place within this class models in which
intersecting D6-branes wrap 3-cycles on toroidal backgrounds~\cite{Blumenhagen:2005mu}, despite the
volumes in this case being restricted to be rather small. The second class of models
are those in which the compact volume is significantly warped and
Randall-Sundrum \cite{Randall:1999vf,Randall:1999ee} like.
In this class of models, which includes the KKLT scenario~\cite{KKLT,Grana:2005jc},
the standard model branes are typically assumed to be located at the bottom of a warped throat.
Hidden branes may be present for a variety of reasons, such
as tadpole and/or anomaly cancellation in the former class, or ``uplift'' in KKLT scenarios.

In the present paper, we shall extend
the discussion of Refs.~\cite{Abel:2003ue,Abel:2006qt} to consider set-ups
in both of these categories, involving
supersymmetric and nonsupersymmetric configurations of D-branes, both
in large volumes and in warped backgrounds with fluxes. Our analysis
(beginning in the following section) will demonstrate that kinetic mixing between
visible and massless hidden $U(1)$s is an interesting possibility to search for
in forthcoming experiments. Clearly the issue of St\"uckelberg masses
and kinetic mixing are related, so one of the main aims of this paper
will be to show how to disentangle them in the string calculation. We will show
using a variety of methods how both the kinetic mixing and
the St\"uckelberg mass mixing can be derived from the underlying type
II string compactification. We will demonstrate by explicit example that
kinetic mixing can occur without St\"uckelberg masses in a completely supersymmetric
set-up where we can use conformal field theory (CFT) techniques. We will then
develop a supergravity approach which allows us to examine the phenomenon
in more general backgrounds, where we
find that kinetic mixing is natural in the context of flux
compactifications.

\subsection{Review: detection of hidden-sector $U(1)$s and current limits}
\begin{figure}
\begin{center}
\epsfig{file=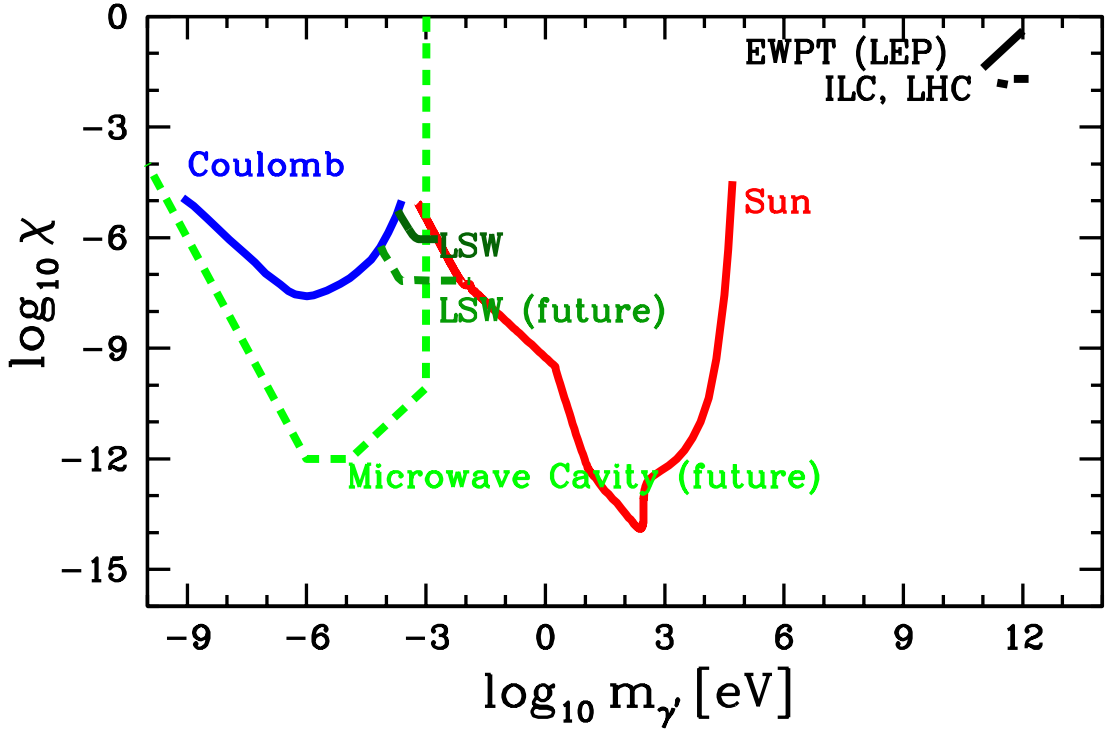,bbllx=20pt,bblly=224pt,bburx=591pt,bbury=605pt,width=15cm}
\end{center}
\caption{Upper limits on the kinetic mixing parameter $\chi$ versus the hidden-sector $U(1)$ gauge boson mass
$m_{\gamma^\prime}$, from electroweak precision tests
(EWPT) at LEP and future experiments at LHC and ILC~\cite{Chang:2006fp,Kumar:2006gm,Feldman:2007wj},
from searches for deviations of the Coulomb law~\cite{Williams:1971ms,Bartlett:1988yy},
and from searches for signatures of
$\gamma\leftrightarrow \gamma^\prime$ oscillations, exploiting, as a photon source,
current and future laboratory lasers (light-shining-through-a-wall (LSW) experiments)~\cite{Ahlers:2007qf},
future microwave cavities~\cite{Jaeckel:2007ch}, or the sun~\cite{Popov:1991,Redondo:2008aa}.    }
\label{kineticmixingpure}
\end{figure}

Before beginning the analysis, we would like to review the possible methods of detection of
hidden-sector $U(1)$s, and the current observational
limits. The masses of the
hidden-sector photons and matter, and the kinetic mixing all come in to play, and
because of this we will here give as general a discussion as possible, in particular
elucidating the experimental differences in the possible detection of massless versus massive hidden $U(1)$s.
Indeed, the best way to search directly for the hidden-sector $U(1)$ gauge boson ($\gamma^\prime$)
depends primarily on its hitherto undetermined mass. For a mass in the range
$m_Z\approx 100\ {\rm GeV} \lesssim
m_{\gamma^\prime}\lesssim 1$~TeV, precision electroweak tests can be used~\cite{Holdom:1990xp,Babu:1997st}
to set an upper limit $\chi\lesssim {\rm few}\times 10^{-2}$ on the mixing parameter
which will be only mildly improved
by future measurements at the high-energy frontier by LHC and
ILC~\cite{Chang:2006fp,Kumar:2006gm,Feldman:2007wj}.
For smaller masses, the best limits arise from searches for
$\gamma\leftrightarrow\gamma^\prime$
oscillations~\cite{Okun:1982xi,Ahlers:2007qf,Jaeckel:2007ch,Popov:1991,Redondo:2008aa}
and for deviations from Coulomb's law
(cf. Fig.~\ref{kineticmixingpure}). Note however that, if the hidden-sector $U(1)$ photons
are massless (i.e. the gauge symmetry is unbroken), then in the absence of light hidden matter
there is no limit on its mixing with hypercharge (because the effect can be reabsorbed by a redefinition of
the hypercharge coupling constant).

\begin{figure}
\begin{center}
\epsfig{file=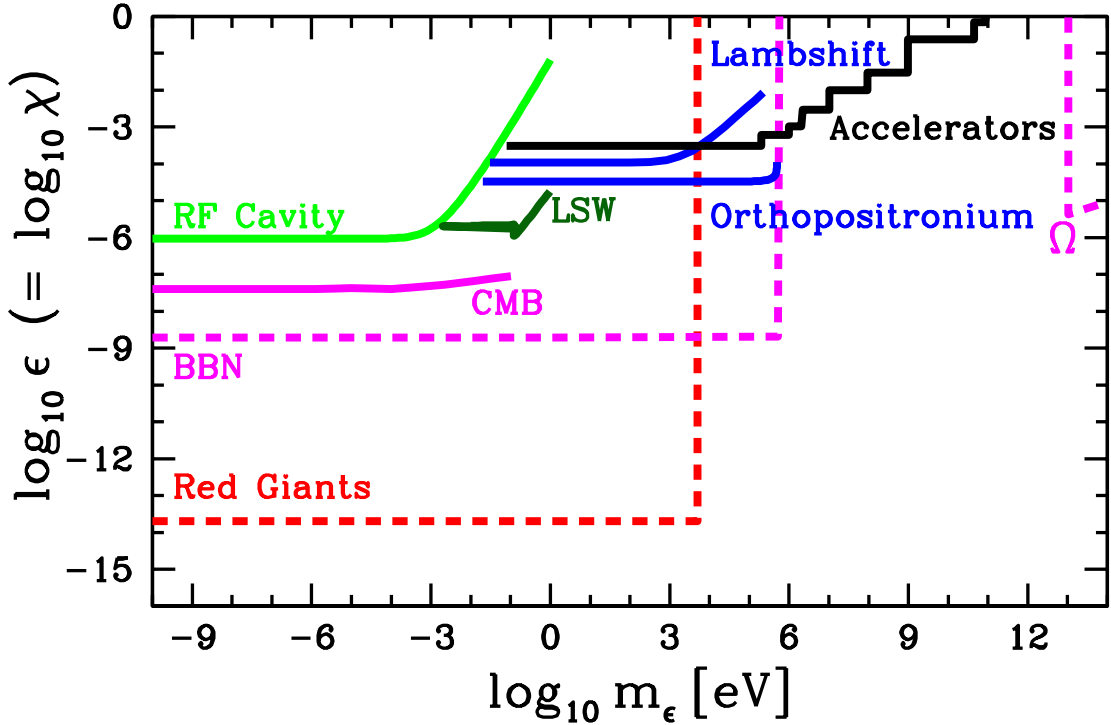,bbllx=20pt,bblly=242pt,bburx=591pt,bbury=605pt,width=15cm}
\end{center}
\caption{Upper limits on the fractional charge $\epsilon = Q_\epsilon/e$ of a hidden-sector
fermion with mass
$m_\epsilon$.
Some of the limits only apply if there is also an ultralight
hidden-sector $U(1)$ gauge boson which gives rise to the minicharge $\epsilon \sim \chi$
by gauge kinetic mixing
with the photon.
Laboratory limits arise from laser polarization and light-shining-through-a-wall (LSW)
experiments~\cite{Ahlers:2007qf}, from energy loss considerations of RF cavities~\cite{Gies:2006hv},
from searches for the invisible decay of orthopositronium~\cite{Gninenko:2006fi}, from
Lamb shift measurements~\cite{Gluck:2007ia} and from searches at
accelerators~\cite{Goldberg:1986nk,Davidson:1991si}.
Limits from cosmology are due the non-observation of a significant distortion of
the spectrum of the cosmic microwave background (CMB) radiation~\cite{Melchiorri:2007sq}
(for a limit exploiting the CMB anisotropy, see Ref.~\cite{Dubovsky:2003yn}),
due to the apparent successfullness of standard big bang nucleosynthesis (BBN)~\cite{Davidson:2000hf},
and due to the observational requirement that the contribution of MCPs to the energy
density should not overclose the universe, $\Omega = \rho/\rho_{\rm crit}<1$~\cite{Davidson:1993sj}.
Finally, an astrophysical limit can be placed by energy loss considerations of
red giants~\cite{Davidson:2000hf}.}
\label{epsilon}
\end{figure}

This is different if, in addition to the possibly light hidden-sector $U(1)$ gauge bosons,
there are light hidden-sector matter particles which are charged under the hidden-sector $U(1)$ gauge symmetry.
These could include for example a hidden-sector fermion $h$ with a bare coupling to $A_\mu^{(b)}$
given by
\beq
\mathcal{L} \supset
\bar{h}\fssd{A}^{(b)}\, h .
\label{bcoupling}
\eeq
Such particles are known to show up as electrically minicharged particles, with their electric charge being
proportional to the gauge kinetic mixing parameter~\cite{Holdom:1985ag}.
Indeed, upon diagonalizing the gauge kinetic term in Eq.~(\ref{LagKM}) by the shift
\begin{equation}
\label{shift}
A_{\mu}^{(b)}\rightarrow \tilde{A}_{\mu}^{(b)}+\chi A_{\mu}^{(a)},
\end{equation}
the coupling term~(\ref{bcoupling}) gives rise to a coupling with the visible
gauge field $A_\mu^{(a)}$,
\begin{equation}
\bar{h}\fssd{A}^{(b)} h\rightarrow \bar{h}\fssd{\tilde{A}}^{(b)} h + \chi \bar{h}\fssd{A}^{(a)}
 h ,
\end{equation}
corresponding to a possibly small, non-integer charge with respect to the
visible sector $U(1)$,
\begin{equation}
\label{epsiloncharge}
Q_h^{(a)}= \chi g_b  \equiv \epsilon\,e .
\end{equation}
Hence in a wide class of models one can also look experimentally for
signatures of the virtual or actual presence of electrically
minicharged particles (MCPs). For low MCP masses,
$m_\epsilon\lesssim 0.1$~eV, the best current laboratory limits on
the electric charge (cf. Fig.~\ref{epsilon}), $\epsilon\lesssim {\rm
few}\times 10^{-7}$~\cite{Ahlers:2007qf}\footnote{If there are
hidden-sector photons in addition to the MCPs this bound may be
somewhat weakened. The most robust bound then comes from
light-shining-through-a-wall experiments discussed
below~\cite{Ahlers:2007rd}.}, are obtained from laser polarization
experiments~\cite{Gies:2006ca,Ahlers:2006iz}, such as
BFRT~\cite{Cameron:1993mr},
PVLAS~\cite{Zavattini:2005tm,Zavattini:2007ee}, and
Q\&A~\cite{Chen:2006cd}, where linearly polarized laser light is
sent through a transverse magnetic field, and changes in the
polarization state are searched for. Such changes would signal that
some photons are being retarded by interactions with virtual
hidden-sector matter (the SM effect being too small to observe), or
are being lost by pair production of hidden-sector matter particles.
Other laser experiments, exploiting a light-shining-through-a-wall
technique, such as ALPS~\cite{Ehret:2007cm},
BFRT~\cite{Ruoso:1992nx}, BMV~\cite{Robilliard:2007bq},
GammeV~\cite{Chou:2007zz}, LIPSS~\cite{Afanasev:2006cv},
OSQAR~\cite{OSQAR}, and PVLAS~\cite{PVLASLSW} are sensitive to
$\gamma\leftrightarrow\gamma^\prime$ oscillations which can be
induced, even for massless $\gamma^\prime$s, by the presence of
virtual hidden-sector matter in a magnetized
vacuum~\cite{Ahlers:2007rd}; current LSW data provide a limit of
$\chi = \epsilon \lesssim 2\times 10^{-6}$, for
$m_{\gamma^\prime}=0$ and $m_\epsilon\lesssim
0.1$~eV~\cite{Ahlers:2007qf}. A comparable laboratory limit,
$\epsilon\lesssim 10^{-6}$, for $m_\epsilon\lesssim 1$~meV, can be
inferred from the non-observation of an excessive energy loss due to
Schwinger pair production of minicharged particles in the strong
electric fields in superconducting accelerator
cavities~\cite{Gies:2006hv}. In the mass range from eV up to the
electron mass, the best laboratory limits, $\epsilon\lesssim 3\times
10^{-5}$, arise from searches for the invisible decay of
orthopositronium~\cite{Gninenko:2006fi}, while in the higher mass
range the accelerator limits dominate; these, however, are rather
loose (cf. Fig.~\ref{epsilon} and
Refs.~\cite{Goldberg:1986nk,Davidson:1991si}). Bounds involving
cosmology or astrophysics are seemingly much better, notably in the
sub-electron mass region (cf. Fig.~\ref{epsilon}). However these
limits, in particular those arising from BBN and energy loss
constraints from red giants, are more model-dependent and can be
considerably milder in certain parameter ranges of hidden-sector
particles and
interactions~\cite{Masso:2006gc,Kim:2007wj,Antoniadis:2007sp}.

It is therefore reasonable to suppose that
current and near future laser experiments have the potential to
detect the presence of a hidden massless or light $U(1)$ gauge field coupled to charged
hidden light ($\lesssim {\cal O}({\rm eV})$) matter. There is nothing to forbid these in the
hidden-sectors of type II string theory; assuming a collection of intersecting branes at
some location of the compact space removed from the visible sector, there will be $U(1)$
factors of various masses and initially massless chiral multiplets.  After supersymmetry breaking
(presumably gravity-mediated) the charged bosons acquire masses, leaving massless fermions that we
shall be interested in probing for; the charged bosons, we shall assume, acquire similar masses to their
visible counterparts and are thus unobservable by the laser experiments previously mentioned.
We shall thus assume no hidden Higgs mechanism acts to give masses to the fermions, although this
could be relaxed provided the masses are sufficiently small.

\subsection{Overview: kinetic versus mass mixing in type II string theory}
\label{sec:Overview}

Before getting to the details of the different scenarios, we should make some
general remarks about kinetic mixing in string theory and outline the various computations
we are going to perform. We will also at this point clarify the interplay between kinetic mixing,
St\"uckelberg mass and anomaly cancellation.

Kinetic mixing can be understood in two ways: either as open strings stretching between separated branes,
or as closed strings propagating between them. In principle, a string CFT computation can give the mixing,
as in the case of toroidal models \cite{Abel:2003ue}.
However that approach is limited in the sense that it can only address what can be done using CFT: it can only
be used in models with backgrounds that are orbifolds or orientifolds of tori.
In order to give a more general discussion it is far more useful to develop the effective supergravity approach.
This reproduces the dominant contributions to kinetic mixing when computed as tree-level closed
string propagation
in toroidal compactifications, and allows us to consider more general
gravitational and/or flux backgrounds, depending on the scenario in question. What the second method sometimes
lacks in rigor, it more than makes up for in generality.

We shall begin our discussion in earnest in the next section by considering kinetic-mixing in the context of
D-brane models in type IIB string theory using the CFT approach.
For example one can think of
supersymmetric models based on an orbifolded torus with an additional orientifolding.
These supersymmetric models, first discussed in Ref.~\cite{Z2Z2Models}, are based on networks of
wrapped intersecting D6-branes.
They are a good starting point because here calculations {\em can} be done using CFT, and
this will help us to develop an intuition for when kinetic mixing will occur and when it will not.
This is a delicate question because  the kinetic-mixing
diagram is also the diagram for the mass term mixing visible and hidden $U(1)$s,
and a single one-loop open string
diagram contributes to both of the terms in the Lagrangian of the form
\begin{equation}
\label{mass}
 m_{ab}^2 A^{(a)}_{\mu} A^{(b)\,\mu} + \frac{\chi_{ab}}{2 g_a g_b} F^{(a)}_{\mu\nu} F^{(b)\, \mu\nu},
\end{equation}
where $m_{ab}$ is the aforementioned St\"uckelberg mass mixing, associated with anomalies and
their cancellation via the Green-Schwarz mechanism
(a discussion of which can be found in Refs.~\cite{Antoniadis:2002cs,Anastasopoulos:2003aj,Kiritsis:2003mc,Anastasopoulos:2004ga,Anastasopoulos:2006cz}).
Massless $U(1)$s must have zero mixing,
$m_{ab}=0$, with all the other $U(1)$s in the
theory\footnote{Note that the converse is not true: absence of 4d anomalies does not guarantee absence
of St\"uckelberg masses because of possible 6d anomalies.
Also mass mixing between U(1)s has been proposed in string models as a means of
supersymmetry-breaking mediation~\cite{Verlinde:2007qk}, but we will not consider this possibility here.}.
Since both of the terms in Eq. \eqref{mass} arise from the same diagrams, how can $\chi_{ab}$ be non-vanishing
between two anomaly-free $U(1)$s when we must also have $m_{ab}=0$?

\begin{figure}
\begin{center}
\includegraphics[bb=0bp 200bp 494bp 823bp,clip,scale=0.35]{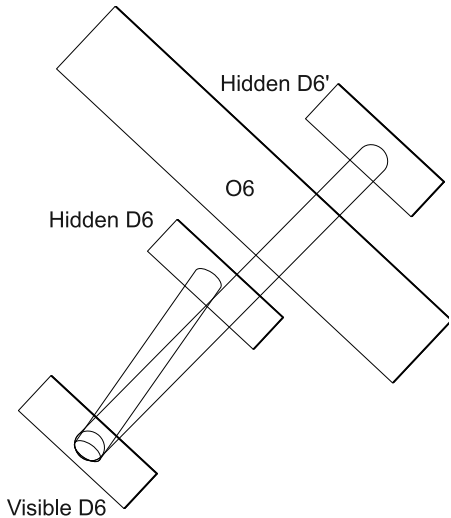}
\end{center}
\caption{Schematic illustration of the reason why kinetic mixing need not cancel between
anomaly-free $U(1)$s. We show contributions to photon mixing with hidden $U(1)$s in the presence
of an orientifold plane: St\"uckelberg mass-terms cancel, whereas kinetic mixing terms do not.}
\label{orientifold-picture}
\end{figure}

The answer is that, in order to
get a contribution to the St\"uckelberg mass, one has to extract a $1/k^2$ pole from the
appropriate one-loop integral (see Eq.~\eqref{V0V0} below).
>From the closed string point of view this corresponds to the St\"uckelberg mass
only getting contributions from {\em massless} closed string modes. Such contributions
are blind to the location in the compact dimensions of the different sources.
The non-pole contributions in this integral give rise to $\chi_{ab}$.
Importantly these contributions to $\chi_{ab}$ are from both massless {\em and} massive Kaluza Klein modes.
The latter certainly do care about the location of the sources in the compact dimensions, and
so contributions to $\chi_{ab}$ do not generally cancel even when the contributions to $m_{ab}$
do\footnote{From a more field theoretic
perspective we can argue as follows. St\"uckelberg masses typically arise from anomalies.
Anomalies, however, do not care
about the masses of the particles, i.e. the length of the stretched open strings. In contrast,
kinetic mixing depends on the masses of the particles going around the loop.}.
A schematic example is shown in Fig.~\ref{orientifold-picture}.
The picture indicates a
localized standard model visible sector (on D6-branes, although the dimensionality is irrelevant)
and a hidden sector $U(1)$ living on a brane together with the image brane in an orientifold plane.
The contributions from the brane to the St\"uckelberg mass mixing between hidden and visible photons cancels
that from its image. The same cancellation does not occur for kinetic mixing, because the hidden brane and its image are separated.

To demonstrate the validity of this general idea, in section~\ref{sec:susy}
we will explicitly compute kinetic mixing
in a supersymmetric and tadpole-free construction on a toroidal background, where
we can calculate it with a straightforward CFT treatment.
If kinetic mixing occurs between anomaly-free $U(1)$s here, then we can be
sure that it can be decoupled from the question of anomaly cancellation. We
begin in section~\ref{sec:CFT} with the generalities of the CFT calculation.
Then, in section~\ref{sec:susy}, we present an anomaly-free toy model similar to those of
Refs.~\cite{Z2Z2Models,Blumenhagen:2005mu}, but of course configured so as to have additional anomaly-free
hidden $U(1)$s.

Following that, in section~\ref{sec:SUGRA}, we will demonstrate that the
same result can be readily computed using the effective supergravity field theory.
As one might expect, the supergravity approach gives a more intuitive and
general understanding which can then be applied to alternative scenarios,
where the global properties of the models are not so well understood.
In section \ref{sec:RS}, we consider a version of the Randall-Sundrum set-up which mimics the effect of warping.
>From this we learn that kinetic mixing can be large in such models and is only
tamed by fluxes generating sufficiently large masses for the mediating closed
string fields. In particular the warping of the metric itself has no effect on the
size of the kinetic mixing. We then confirm this in section \ref{sec:KT}
by considering the more stringy set-up of $U(1)$s located at the tip of a Klebanov-Tseytlin throat.

\section{The CFT computation of kinetic mixing: generalities}\label{sec:CFT}

We begin by reconsidering kinetic mixing in flat backgrounds where we can use CFT.
Technically the computation is identical to finding gauge threshold corrections
(cf. also Refs.~\cite{Lust:2003ky,Berg:2004ek,Blumenhagen:2006ux}),
but with a trivial but crucial difference: the group-theoretical prefactors are changed.
Typically, an anomaly-free $U(1)$ is composed of a
linear combination of the $U(1)$s coming from different stacks of branes.
The different $U(1)$s will be labelled $a,b$ and the stacks will get labels $i,j$\footnote{When orientifolds
are present the images are counted as different stacks.}.
The vertex operator describing $U(1)^{a}$ is given by
\begin{equation}
\label{vertexop}
V^{a}=\sum_{i} c^{a}_{i} V^{a}_{i},
\end{equation}
where we sum over stacks $i$ and the constants $c^{a}_{i}$ are chosen so that the corresponding
$U(1)$ is anomaly free.  The individual vertex operator on a given stack of branes $i$ is given by
(in the zero-picture)
\beq
V^a_{i\, } = \lambda_i^a \varepsilon_{\mu} (\partial X^{\mu} + 2\ap (k \cdot \psi) \psi^{\mu})
e^{\imi k \cdot X},
\eeq
where as usual $\varepsilon_\mu$ is the polarization vector, $\psi^\mu$ and $X^\mu$
are worldsheet fermions and bosons respectively and
$\lambda^{a}_{i}$ is the Chan-Paton matrix on the stack $i$. The anomaly free $U(1)$s
are the linear  combinations that obey
\beq
\sum_i  c^a_i \tr_i (\lambda_i^a) = 0\, .
\eeq
The general expression for the amplitude we calculate is then
(in the closed string channel)~\cite{Antoniadis:2002cs,Abel:2003ue}
\begin{multline}
\langle V^a_i V^b_j \rangle = 4 (\ap)^2 \tr_i (\lambda_i^a) \tr_j (\lambda_j^b)
\varepsilon_\mu \varepsilon_\nu
( g^{\mu \nu} k^2 - k^{\mu} k^{\nu} ) \int_0^{\infty} dl \int_0^1 dx\, e^{k^\mu k^\nu G_{\mu \nu}} \\
\sum_{\nu} \bigg[\frac{\theta_4^{\prime \prime} (x)}{\theta_4(x)} -
\frac{\theta_{\nu}^{\prime \prime} (0)}{\theta_{\nu}(0)}\ \bigg] \frac{1}{(8\pi^2 \ap)^2}
\frac{\theta_{\nu} (0)}{\eta^3 (\imi l)} Z_{\nu}^{ij} (\imi l)\, ,
\label{V0V0}\end{multline}
 where the Green function is given by
\begin{equation}
G_{\mu \nu} (x) = -2 \ap g_{\mu \nu} \log \left| \frac{1}{l} \frac{\theta_4 (x)}{\eta^3 (\imi l)} \right| \, ,
\end{equation}
$\theta$ and $\eta$ are the elliptic theta and Dedekind eta functions,  and
$\tr_i, \tr_j$ denote that the trace is to be taken over the individual branes inside a stack $i$ or $j$.
$Z_\nu^{ij}$ is the
partition function with spin structure $\nu$, with the
${ij}$ indices indicating that it is in general a function of the
displacements between the stacks.  In the open string channel, this is a non-planar diagram (i.e. with
the two vertex operators placed on different boundary stacks).

The final result for kinetic mixing between say hypercharge $A_\mu^a$ and a hidden anomaly-free $U(1)$ $A_\mu^b$
contains a further $i,j$ sum over all the relevant stacks
contributing to each $U(1)$ as dictated by Eq.~\eqref{vertexop}.
Note that if we are considering an orbifold model with fractional branes where the branes are separated,
only the non-fractional component contributes: there can be no contribution from twisted sectors because
a displacement between the ends of the annulus diagram is not consistent with an orbifold twist.

The form for the amplitude in the on-shell/low-energy limit $k^2 \rightarrow 0$ will then be
\begin{equation}
\langle V^a V^b \rangle = m^2_{ab} A^a_{\mu} A_b^{\mu} +
\frac{\chi_{ab}}{ g_a g_b} k^2 A^a_{\mu} A_b^{\mu}\, .
\end{equation}
Since the right hand side of Eq.~\eqref{V0V0} explicitly contains the transverse structure \linebreak
\mbox{$k^2g^{\mu\nu}-k^{\mu}k^{\nu}$}, the contribution to the mass must
come from a $1/k^2$ pole of the integral. To make the pole structure manifest we take the large $l$ limit,
\beq
e^{k^\mu k^\nu G_{\mu\nu}} = e^{-\frac{1}{4}2\pi\alpha' k^2 l },
\eeq
while the rest of the integrand can be generically expanded as
\beq
\sum_{\nu} \bigg[\frac{\theta_4^{\prime \prime} (x)}{\theta_4(x)} -
\frac{\theta_{\nu}^{\prime \prime} (0)}{\theta_{\nu}(0)} \bigg]
\frac{1}{(8\pi^2 \ap)^2} \frac{\theta_{\nu} (0)}{\eta^3 (\imi l)} Z_{\nu}^{ij}
(\imi l) \propto 1 + \sum_{\beta_{ij}>0} N(\beta_{ij}) e^{-\pi \beta_{ij} l},
\eeq
where $N(\beta)$ counts the multiplicity of closed-string modes at level $\beta$ (which
includes Kaluza-Klein and winding modes). The
first term corresponds to massless closed string states and results in a $1/k^2$ pole, generating a
mass-like term for the gauge fields.
This term is $i,j$ independent.
On the other hand, $\chi_{ab}$ receives contributions from the second term which does depend on the
displacement between $i$ and $j$.

As we have already stated, the sum of all contributions to the mass-term
must be absent between two anomaly-free gauge groups. These have been calculated in, for example,
Ref.~\cite{Antoniadis:2002cs},
where the role of the mass-mixing term is elucidated as the St\"uckelberg mass generated for
anomalous gauge groups, which
emerges when the gauge field ``eats'' the relevant closed string modes.
(In the case of $\dthree$-branes the modes
that are eaten are NS-NS $B_2$ fields and R-R $C_2$ fields, as will become apparent later.)
At the same time, the kinetic mixing parameter $\chi$ gets contributions from both the pole {\em and}
the non-pole parts of the integrand, with the latter corresponding to massive intermediate states with
$\beta\neq 0$.

To give a specific illustration, consider the example of a single D3 brane and a $\dthreebar$ brane
on a $T^6$ factorized into 3 complex 2-tori labelled by $\kappa=1\ldots 3$:
\begin{equation}
 Z_{\nu}^{ij} (\imi l) = \frac{1}{2}\delta_{\nu}^{\prime}
 \frac{(\ap)^3 (2\pi)^6}{8V_6}
 \frac{\theta_{\nu}^3 (0)}{\eta^9 (\imi l)} \prod_{\kappa}
 \sum_{q^{\kappa},p^{\kappa}} \exp \bigg[ - \frac{\pi \ap l}{2 T_2^{\kappa} U_2^{\kappa}} | q^{\kappa}
 + \ov{U}^{\kappa} p^{\kappa}|^2 -\frac{2\pi \imi}{U_2^{\kappa}} {\rm Im}(z_{ij}^{\kappa}
 (\ov{U}^{\kappa} p^{\kappa} + q^{\kappa}) ) \bigg]
\end{equation}
where $U^{\kappa}, T^{\kappa}$ are the complex and K\"ahler moduli, and $z_{ij}^{\kappa}$ is
the complex separation vector between branes $i$ and $j$ (scaled to be dimensionless;
the true distance is $2\pi z_{ij}^{\kappa} \sqrt{T_2^{\kappa}/U_2^{\kappa}}$).
The above contains the expected massless mode whose effect can be seen in the $l\rightarrow \infty$ limit:
\begin{multline}
\langle V^a_i V^b_j \rangle =  \tr_i (\lambda_i^a) \tr_j (\lambda_j^b) \varepsilon_\mu \varepsilon_\nu
( g^{\mu \nu} k^2 - k^{\mu} k^{\nu} ) \int_0^{\infty} dl \frac{(2\pi \ap)^4}{4\ap V_6} e^{-\frac{1}{4} 2\pi
\ap k^2 l} \\
\bigg\{ 1 + \prod_{\kappa} \sum_{q^{\kappa},p^{\kappa} \ne 0}
\exp \bigg[  - \frac{\pi \ap l}{2 T_2^{\kappa} U_2^{\kappa}} | q^{\kappa} + \ov{U}^{\kappa} p^{\kappa}|^2
-\frac{2\pi \imi}{U_2^{\kappa}} {\rm Im}(z_{ij}^{\kappa} (\ov{U}^{\kappa} p^{\kappa} + q^{\kappa}) ) \bigg]
\bigg\}
( 1 + \mathrm{string\  mass\  terms}) .
\label{ViaVjb}
\end{multline}
We can see explicitly that only the first term, $1$, in the curly brackets gives a $z_{ij}$ independent
contribution.
The second term depends
explicitly on the positions $z_{ij}$. Since it corresponds to \emph{massive} closed string modes
it contributes only to the kinetic mixing.

The reason for neglecting terms of order the string mass in \eqref{ViaVjb} is the familiar exponential damping
of massive modes beyond their wavelength; the phase factor in the above ensures that it is an accurate
approximation to consider only the Kaluza-Klein expansion of the tori. This implies that it is the closed
string modes which heavily dominate the process once the branes are separated by more than a string length,
and moreover that we should be able very accurately to reproduce the expression above using only field theory
as we shall do in section~\ref{sec:SUGRA}.

The expected mass-mixing between the $U(1)_a$ of the D3, and the $U(1)_b$ of the $\dthreebar$
coming from the massless modes is found to be
\begin{equation}
\mathcal{S} \supset  \int d^4 x  \frac{1}{\ap} \frac{(2\pi \ap)^3}{V_6} A_{\mu}^a A^{b \, \mu}\, .
\label{m2}\end{equation}
Note that  in the D3-\dbar{3} system we also have a contribution from planar diagrams (i.e. with both vertex
operators placed on the same boundary in contrast with Eqs.~(\ref{V0V0}),(\ref{ViaVjb})).
This generates gauge threshold corrections but more importantly
renders any gauge group carried by the brane or antibrane massive. This is a consequence of the uncancelled
NS-NS charges, i.e. that there is a nonzero cosmological constant. Hence, due to the volume suppression of
the masses, this could be a candidate for the St\"uckelberg $Z'$ scenario; we make some remarks on this in
appendix \ref{D3D3BRemarks}. Moreover there is kinetic mixing. We obtain for $\chi$ in Eq.~\eqref{LagKM}
\begin{equation}
\chi_{ab} \approx g_a g_b  \frac{(2\pi \ap)^3}{V_6}
\sum_{q^{\kappa},p^{\kappa} \ne 0} \frac{\exp\left[ \sum_{\kappa}  -
\frac{2\pi \imi}{U_2^{\kappa}} {\rm Im}(p^{\kappa} z^{\kappa}
\ov{U}^{\kappa} + q^{\kappa} z^{\kappa}) \right]}{\sum_{\kappa}
\frac{\ap }{T_2^{\kappa}U_2^{\kappa}} | q^{\kappa} + \ov{U}^{\kappa}
p^{\kappa}|^2} \label{D3D3barmixing}\end{equation} where $z^k$ is
the displacement between the brane and anti-brane in the $k$'th
complex 2-torus. Note that the kinetic-mixing term produced by the
string amplitude, $\chi_{ab}/g_ag_b$, actually contains no factors
of the gauge coupling (since the vertex operators carry none), and
that in addition $\chi$ depends on $z^k$ but the mass-mixing does
not.

\subsection{When can we have kinetic mixing between massless $U(1)$s?}
\label{sec:2.1}

We now wish to show that kinetic-mixing can occur between anomaly-free $U(1)$s.
To begin with, note that the amplitude is always proportional to the trace of the Chan-Paton factor for the $U(1)^a$
gauge factor $\tr_i (\lambda_i^a)$. This is also the factor for the mass-mixing term, and so if we wish to avoid a
massive gauge field, the total contribution for gauge group $a$ must vanish: if the $U(1)$ charge is given by a linear
combination of the trace $U(1)$ charges on the branes as
$
Q^a=\sum c^a_i Q^a_i\, ,
$
then we must have either
\beq
\sum_i  c^a_i \tr_i (\lambda_i^a) = 0\, ,
\eeq
or its equivalent for the hidden sector $U(1)$s. However if we are considering a $U(1)^a$ split among
separate branes, we cannot simply factor out $\sum_i c^a_i \tr_i (\lambda^a_i) $ from the
kinetic-mixing term because the integrand depends on the positions (on $z_{ij}^k$ in other words).

As an example
if we have a $U(1)_a$ split among separate branes, this can mix with other $U(1)$ fields.
This can also be understood in field theory as the embedding of the $U(1)$ as the generator of a broken
non-Abelian group;
for example for two branes carrying naively a $U(2)$ when coincident, upon splitting there are two $U(1)$
charges $Q_1$
and $Q_2$. The combination $\frac{Q_1}{2} + \frac{Q_2}{2}$ is anomalous, and corresponds to the $U(1)$
in $U(2)=U(1)\times SU(2)$, while the combination $\frac{Q_1}{2} - \frac{Q_2}{2}$ is always non-anomalous and
corresponds to the $\sigma_3$ generator in $SU(2)$.

If there is an orientifolding (which is generally required in order to cancel tadpoles), then it is
typically accompanied by a reflection $R$ on the compactified coordinates  which generates orientifold planes
and D-brane images.
Then we automatically have pairs of branes.
 Under the orientifold action, each brane $i$
has an image $i'$, which carries the same gauge group. The Chan-Paton matrices of the brane and its image are
related by
\beq
\lambda_{i}'=\mp \gamma_{\Omega R}^{-1} \lambda_i^T \gamma_{\Omega R}\,,
\eeq
where $\gamma_{\Omega R}$ determines the action of the orientifold on the Chan-Paton matrices, and where the
minus (plus) sign is fixed by consistency of the model,
and arises from the odd number of worldsheet fermions in physical states imposed by the GSO projection.
 Since the relevant charges come from $\tr(\lambda_i)$, however, the image branes have charges
 $Q_{i'} = \mp Q_i$. For such branes sitting on the orientifold we find a $USp(2N) (SO(2N))$ gauge group,
 for $N$ branes and $N$ images, and note that the trace generator is in the former case projected out.
 When the branes are separated from the orientifold plane, however, while remaining in the same
 homology class\footnote{Such as where there are translation or Wilson line moduli, or in a separate
 region of a Calabi-Yau related to the original by an involution of the compact space.
 It is important that the quantum states be related, otherwise the contributions to the gauge
 field mass between brane and image will not be equal.} we automatically have a non-anomalous $U(1)$.
 Again this non-anomalous $U(1)$
can be thought of as coming from a traceless generator of the symplectic group when the branes are on
top of the orientifold plane.  Indeed, in the presence of an orientifold plane, the mass-mixing between
the $U(1)$$_i$ coming from the hidden brane gets contributions from the brane and its image, which together
will be proportional to
$\tr_i (\lambda_i) +\tr_i (\lambda_i') = 0\, $. Notice that this is independent of whether the
visible $U(1)$ in question
is itself anomalous: the $U(1)$ from an isolated D-brane does not get masses from any source if it is
parallel to an orientifold plane, but since the brane and its image are displaced it can still kinetically
mix. In fact, in a large volume compactification, the orientifold and the image can be
removed to large distances and the resulting kinetic-mixing would be dominated by the single brane.
In the closed string picture, the orientifolding projects out the massless modes that transmit the mass-mixing.
A general illustration of this scenario is given in
figure~\ref{orientifold-picture}.

\section{Supersymmetric models}\label{sec:susy}

In order to confirm that kinetic mixing can indeed occur between
anomaly free and massless $U(1)$s
we will now, as promised in the
Introduction,
examine self-consistent (i.e. tadpole-free)
global configurations that have non-vanishing
kinetic mixing between mutually supersymmetric branes.

A convenient framework in which to construct
supersymmetric models consists of a simple orientifold with D6 branes
and O6 planes in type IIA string theory, as reviewed in
Ref.~\cite{Blumenhagen:2005mu}.
In principle, one would like to construct an ${\cal N}=1$ model
similar to those of Ref.~\cite{Z2Z2Models}, but with hidden $U(1)$s.
However the D6-branes wrap all the internal cycles
and typically they always intersect.
Hence it is difficult to construct  ${\cal N}=1$  models
with {\em any} hidden sector.
Our main aim here is not to construct a realistic model, but rather
to provide a simple proof of concept in a completely supersymmetric
set-up, for which ${\cal N}=2$ models will be sufficient.
Such models correspond to dimensionally reduced 6-dimensional ${\cal N}=1$
models, and so St\"uckelberg masses, if they are present,
correspond to the Green-Schwarz anomaly cancellation
of 6-dimensional anomalies (the 4-dimensional anomalies in ${\cal N}=2$ being of
course zero).

\begin{figure}
\begin{center}
\includegraphics[trim = 10mm 175mm 10mm 15mm,clip,width=150mm]{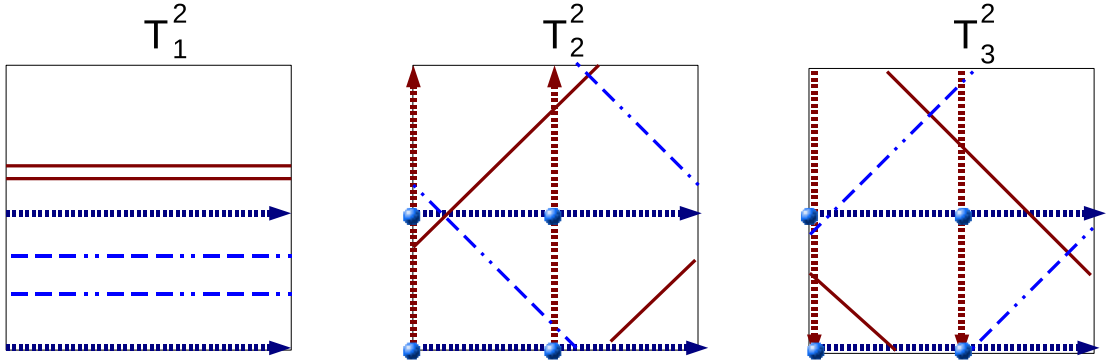}
\end{center}
\caption{Supersymmetric configuration corresponding to the model in Table~1.
Solid lines denote $A$ stacks and dashed-dotted lines represent $B$
stacks. Each of these stacks is separated into $A_1$, $A_2$, and $B_1$, $B_2$ in the first
torus only. The orientifold planes are represented by the dashed lines with arrows. In the first torus,
the two sets of orientifold planes are coincident. Finally the dots on each of
the last two tori show the orbifold fixed points (or more precisely planes). }
\label{susy-config}
\end{figure}

Our general configuration is as follows: we choose
the ten-dimensional spacetime to be $\mathbb{R}^{3,1}
\times \mathbb{T}^2 \times (\mathbb{T}^2\times\mathbb{T}^2)/{\mathbb{Z}_2}$,
where the ${\mathbb{Z}_2}$ orbifolding is taken to act on the
second and third $\mathbb{T}^2$ tori. The tori are all taken to be
rectangular. Denoting the complex coordinates on the compact space
$z^i \in \mathbb{T}^2_i$, the orbifold involution acts as
$\theta: (z_2,z_3) \rightarrow (-z_2,-z_3)$.
The orientifold involution is then introduced as follows. It consists of world
sheet parity transformation $\Omega$ coupled with a non-holomorphic
reflection $R$ in the internal complex cooordinates,
$R:z^k\rightarrow \overline{z}^k$.
The projections leave $ 4 \times 4=16$ fixed points of the $ \mathbb{Z}_2$ orbifold
(see Fig.~\ref{susy-config}), and
16 orientifold fixed planes (O6-planes), 8 for each
of the orientifold actions, $\Omega R $,  $\Omega R \theta$.

This orbifold is a singular limit
of the Calabi-Yau manifold $\mathbb{T}^2 \times K3$.
Although we can calculate
exactly only in the orbifold limit, we can make
completely general statements about how the D6-branes should wrap
in order to obtain kinetic mixing:
we must place our branes on bulk rather than exceptional
cycles; the D6 branes must wrap two-cycles on the $K3$, and a one-cycle on
the $\mathbb{T}^2$; thus both the closed and open string sectors
preserve ${\cal N}=2$ supersymmetry in four dimensions (such as prior
to $D$-term SUSY-breaking in split-supersymmetry models
\cite{Antoniadis:2006eb})\footnote{Despite the fact that the kinetic mixing can be calculated exactly only
in the orbifold limit case, these statements are valid more generally
because the calculation of mixing in ${\cal N}=2$ depends only on the
zero modes, with the massive string excitations cancelling
\cite{Dixon:1990pc,Antoniadis:1996vw} - and thus depends only on the
intersection form on the $K3$.}.

In the calculable orbifold case,
we may derive rather general expressions for the
subsequent kinetic mixing before presenting an explicit model
(detailed in Table~\ref {WrapSimple} and Fig.~\ref{susy-config}).
We need only assume that the two massless gauge
groups $U(1)_a$ and $U(1)_b$ come from two
parallel stacks of branes each, labelled $A_1,A_2$ and $B_1,B_2$.  In
order to not intersect, they must be parallel to the orientifold plane
in torus 1, but not lie upon it (cf. Fig.~\ref{susy-config}). We denote the separations from the
O6-plane in the torus 1 $y_{\sst A_i}$, and write
$\delta_{ij} \equiv y_{\sst A_i} - y_{\sst B_j}$.
Note that it is possible to take $A_2 = A_1'$ (or similarly
$B_2 = B_1'$), the image under the orientifold, but we shall not
require this. The charges for the massless combinations are given by
$Q_a = \frac{1}{N_{A_1}} Q_{A_1} - \frac{1}{N_{A_2}} Q_{A_2} = \sum_i \frac{c_i^a}{N_{A_i}} Q_{A_i} $,
and similarly for
$Q_b$, where $N_{A_i}$ is the number of branes in
stack $A_i$, and $Q_{A_i},Q_{B_i} = \pm 1$.
The kinetic mixing is then given by
\beq
\chi = \sum_{ij} c_i^a c_j^b Q_{A_i} Q_{B_j} \chi_{ij} =
\chi_{11} - \chi_{12} - \chi_{21} + \chi_{22},
\eeq
where \cite{Berg:2004ek,Abel:2006yk}
\beq
\chi_{ij} = \frac{g_a g_b}{4\pi^2} I_{AB} \bigg[\log \left|
\frac{\theta_1 (\frac{\imi \delta_{ij} L_1}{2\pi^2 \ap},
\frac{\imi T^2_1}{\ap})}{\eta(\frac{\imi T^2_1}{\ap})} \right|^2 - \frac{\delta_{ij}^2}{2\pi^3 \ap}
\frac{(L_1)^2}{T_2^1}\bigg]\, ,
\label{OneParallel}\eeq
where $\chi_{ij}$ is the kinetic mixing between $A_i$ and $B_j$;
and $I_{AB}$ is the number of intersections between
the branes in the non-parallel directions, $L_1$ is the length of both branes
in the torus 1 in which they are parallel, and finally
$T_2^1$ is the K\"ahler modulus of torus 1, proportional to the product of the radii
in the case of rectangular tori.
Note that the above can also be calculated \emph{exactly} by the
effective supergravity techniques that will be introduced
in the next section,
since supersymmetry ensures that all of the string mass excitations do not
contribute. Note also that it is crucial that the two stacks of branes
preserve a mutual ${\cal N}=2$ supersymmetry; if they only preserved
${\cal N}=1$ supersymmetry there would be no dependence on the
separation, and thus we could not separate mass mixing from kinetic
mixing, and if they preserved ${\cal N}=4$ the amplitude would
cancel. This is, however, merely a peculiarity of the very symmetric
toroidal orientifold setup that we are using, and the fact that we are
using D6-branes.

\begin{table}
$$
\renewcommand{\arraystretch}{1.5}
\begin{array}{|c||c|c|c||} \hline
   \mbox{stack} & \mbox{$N$} & \mbox{gauge group} & (n^1,m^1)\times (n^2,m^2) \times (n^3,m^3) \\ \hline
   A & 4+4 & SU(2)\times SU(2) ,Q_{A_1},Q_{A_2}  &(1,0) \times (1,1) \times (1,-1) \\
   B & 4+4 & SU(2)\times SU(2) ,Q_{B_1},Q_{B_2} & (1,0) \times (1,-1) \times (1,1)  \\ \hline
\end{array}
$$
\caption{\em Wrapping numbers for a very simple model with kinetic mixing.
As usual, $N$ counts branes plus their
orbifold images.}\label{WrapSimple}
\end{table}

It is easy to see that for generic values of the brane positions,
and in the presence of massless fermions, the induced mixing
would violate the bounds on
kinetic mixing by many orders of magnitude. This is because, with wrapped D6-branes,
one is unable to take a large volume limit to try and dilute it, and in addition (when the cycles
wrapped by the branes and the bulk radii are all of the same order) dilution only occurs for $p\leq 5$.
It is possible to give a mixing of the order
of $10^{-6}$ or less by tuning the configuration: suppose that
$(y_{\sst B_1} - y_{\sst B_2}) \sim T^2_1/L_1 \gg l_s > (y_{\sst A_1} - y_{\sst A_2})$, i.e. the branes are placed generically but one
splitting is much larger than the other, ensuring $\delta_{ij} > l_s$ (since for small $\delta_{ij}$
the mixing grows logarithmically). We then find
\beq
\chi =  \frac{ g_a g_b }{4\pi^2} I_{AB} \bigg[ \frac{(y_{\sst A_1} - y_{\sst A_2})(y_{\sst B_1} - y_{\sst B_2})}{\pi^3 \ap}
\frac{(L_1)^2}{T^2_1} \bigg]\, + O(e^{-\pi T^2_1/\ap}).
\label{chiapprox}\eeq
and thus $ (y_{\sst A_1} - y_{\sst A_2}) \sim 10^{-6} \frac{l^2_s}{L_1}$.
Note that despite the fact that
this distance is much smaller than the string scale, the expressions are still valid since they are
derived from the complete CFT, and moreover such displacements are quite natural when
considered from the field theory perspective since they represent
a Higgsing of the gauge group by giving a vacuum expectation value to an adjoint scalar.

As an extremely simple, explicit, example, we present the
wrapping numbers for two stacks of branes in table
\ref{WrapSimple}. The configuration including orientifold
planes is shown in Fig.~\ref{susy-config}.

The model has to satisfy a number of consistency conditions.
First to preserve supersymmetry, the radii of the
tori are constrained. Denoting by $\theta_i$ the
angle in the $i^{\mbox{\scriptsize{th}}}$ torus between the branes and the O6$_{\Omega R}$-planes,
we must have $\theta_1+\theta_2+\theta_3=0$; in the present ${\cal N}=2$ case we have
$\theta_1=0$, which leads to
\beq
U^{(2)}_2 \frac{m_2}{n_2} +  U^{(3)}_2 \frac{m_3}{n_3} =  0,
\eeq
where in the rectangular case the complex structure moduli $U^{(i)} = \imi U^{(i)}_2 = \imi R_2^{(i)}/R_1^{(i)}$ are simply
the ratio of torus radii. This condition may trivially be satisfied provided $n_i/m_i$ is the same for
all the branes
up to an overall factor.
The tadpole cancellation conditions that must be satisfied are as follows. First
we ensure that the R-R (7-form) charge contribution
from the orientifold planes cancels that of the D6-branes.
The homology class of a brane $A$ with wrappings $(n^i,m^i)$ (where $i$ labels the
tori) is $[\Pi_A] =  \sum_{i=1}^3 n^i [a_i] + m^i [b_i]$,
where the canonical $[a_i]$ cycles correspond to the ${\rm Re}(z_i)$ coordinate,
and the $[b_i]$ cycles to ${\rm Im}(z_i)$. The images under the orientifold have
homology $[\Pi_{A'}] =  \sum_{i=1}^3 n^i [a_i] - m^i [b_i]$. The O6-planes
corresponding to $\Omega R$ (henceforth denoted O6$_{\Omega R}$), have wrapping numbers
$(1,0)\times (1,0)\times (1,0)$, and
hence homology $[\Pi_{O_{\Omega R}}]=[a_1] \times[a_2]\times  [a_3] $, while
those corresponding to $\Omega R \theta$ (henceforth O6$_{\Omega R\theta }$), have wrapping numbers
$(1,0)\times (0,1)\times (0,-1)$, and
homology $[\Pi_{O_{\Omega R \theta}}]=-[a_1] \times[b_2]\times  [b_3] $.
Their D6 charges are $-4$, so the tadpole cancellation
condition is
\begin{equation}
\sum_A N_A ([\Pi_A]+[\Pi_{A'}])=4\times 8 \, ([\Pi_{O_{\Omega R}}]+[\Pi_{O_{\Omega R \theta}}])\, ,
\end{equation}
which, assuming all the branes have $m^1=0$, yields only two constraints on the
wrapping numbers:
\begin{eqnarray}
\sum_A N_A n_A^1n_A^2n_A^3&=&16, \nonumber \\
\sum_A N_A n_A^1m_A^2m_A^3&=&-16 \, .
\end{eqnarray}
These constraints are clearly satisfied by the model in table
\ref{WrapSimple}. (Supersymmetry then ensures the cancellation of the NS-NS tadpoles.)

Kinetic mixing arises when the stacks of branes are
displaced from the O6$_{\Omega R}$-planes in
the first torus. The counting goes as follows. Begin with $N$ branes plus their orientifold images
on top of the orientifold plane,  and passing through orbifold fixed points.
As we move away from the fixed points/planes we get images under
both the orbifold and the orientifold, so that we have a $U(N/2)$ gauge
group. By further splitting the stacks in the first torus, we obtain two separate
$U(N/4)$ gauge groups, and by taking the trace generator from each, we can form
massless $U(1)$ combinations $Q_a = \frac{4}{N_A} (Q_{A_1} - Q_{A_2}),
Q_b = \frac{4}{N_B} (Q_{B_1} - Q_{B_2})$
as described in section~\ref{sec:2.1}
(where 4 counts orbifold images).
Note that St\"uckelberg masses arise for the orthogonal $U(1)$ combinations,
 $Q_{\bar{a}} = \frac{4}{N_A} (Q_{A_1} + Q_{A_2})$ and
$Q_{\bar{b}} = \frac{4}{N_B} (Q_{B_1} + Q_{B_2})$, as expected.

The kinetic mixing for this particular model is given by $2\chi$, with
$\chi$ as in equation (\ref{chiapprox}); note that there is no mixing
between the branes and orientifold images, since $B'$ is parallel to
$A$ (and thus preserves a mutual ${\cal N}=4$ supersymmetry, cancelling any
mixing), but the overall factor of $2$ accounts for $\chi_{A_1' B_1'}
= \chi _{A_1 B_1}$ etc. The above expression will be non-zero provided
that the $y$'s are not equal.

We can straightforwardly find more realistic (although still ${\cal N}=2$) models, in particular ones that have
massless hypermultiplets (since this is what would be required to detect the kinetic mixing).
A tentative model is given in table \ref{Wrap}, which contains the standard-model like
group factors. Again the branes must be separated from the orbifold fixed points and orientifold planes.
This time, the separations in torus one must be such that $y_{\sst A_1} = y_{\sst B_1}=y_{\sst C_1}$, and
$y_{\sst C_2}=y_{D}$.
This ensures that there is (non-chiral) matter charged under the visible gauge groups, and also some charged
(only) under the hidden gauge group $Q_h$. Note that stack $D$ is split to $U(1)_h$ via two $U(2)$s in the
second and third tori; alternatively it could remain as a stack of two branes and two images, giving a
hidden massless $U(1)$ and $SU(2)$. Once more there is kinetic mixing between massless $U(1)$s;
the ``hypercharge'' is given by
\beq
Q_Y = \frac{1}{3} Q_{A_1} - Q_{A_2} + Q_{C_1} .
\eeq
Note that, since branes $C_i$ are parallel to the orientifold plane, they automatically carry massless
gauge groups, and also participate in kinetic mixing. (Because this is still an ${\cal N}=2$ model,
we will not go on to present the spectrum here.)

\begin{table}
$$
\renewcommand{\arraystretch}{1.5}
\begin{array}{|c||c|c|c||} \hline
   \mbox{stack} & \mbox{$N$} & \mbox{gauge group} & (n^1,m^1)\times (n^2,m^2) \times (n^3,m^3) \\ \hline

   A_1 & 6 & SU(3),Q_{A_1} &(1,0) \times (1,1) \times (1,-1) \\
   A_2 & 2 & Q_{A_2} &(1,0) \times (1,1) \times (1,-1) \\
   B_1 & 4 & SU(2) & (1,0) \times (0,1) \times (0,-1)  \\
   C_1 & 2 & Q_{C_1} & (1,0) \times (1,0) \times (1,0) \\ \hline
   D   & 4 & Q_h  & (1,0) \times (1,-1) \times (1,1) \\
   C_2 & 2 & Q_{C_2} & (1,0) \times (1,0) \times (1,0) \\ \hline
\end{array}
$$
\caption{\em Wrapping numbers for a slightly more realistic model.}\label{Wrap}
\end{table}


The above discussion is of course for a very simple model, and we would of course like to build more
realistic examples with
${\cal N}=1$ supersymmetry that can then be broken, and genuine chiral matter charged under the correct
gauge groups etc.
However, if we attempt to realise such models on a torus with an unresolved orbifold we encounter some
obstructions.
Since we require the cancellation of masses but not of kinetic mixing, we need branes with bulk
components that are separated
between the hidden and visible $U(1)$ factors in order that they are hidden, and also split into stacks
so that they are massless.
Thus, given that the mixing in these models is generically large, and that it is difficult
(although not impossible)
to obtain a truly hidden sector, such models are not of particular interest. More importantly, however,
in the current paradigm of LARGE volume \cite{Balasubramanian:2005zx,Conlon:2005ki,Conlon:2006wz}
or KKLT \cite{KKLT} models we should consider a collection of D6-branes or their T-dual in terms of
D3 and D7 branes with gauge fluxes realising the standard model gauge group and spectrum to be a
mere \emph{local} construction supported in some small region of a larger manifold.
It is from this
perspective that we see that a truly hidden sector separated from the visible one by several string
lengths is entirely natural.

Of particular interest are models involving (anti) D3-branes that move in the bulk.
These are required for example in the KKLT scenario to uplift to a de Sitter vacuum~\cite{KKLT,Grana:2005jc},
and may even play the role of an inflaton.
Since the charges of the D3-branes can be cancelled by O3-planes (which may be well separated from them),
fluxes or D7-branes wrapping cycles with non-trivial curvature, there
should be no reason that they may not exhibit kinetic mixing.
If the D3-branes and the branes supporting hypercharge are located at generic positions in some Calabi-Yau,
the nett effect would be one of volume suppression similar to the flat space case \cite{Abel:2006qt}. However
in many scenarios, such as KKLT, the hypercharge is placed at a special position, for example the
tip of a warped throat. One expects that this could drastically alter the phenomenon of
kinetic mixing: not only is the background now warped, but the fluxes that cause the
warping also give masses to the very fields that mediate the kinetic mixing.
In order to analyse these more general cases, we shall have to go beyond the flat space
approximation and develop a supergravity approach.


\section{The supergravity calculation of kinetic-mixing}\label{sec:SUGRA}

As a warm-up exercise for the supergravity approach, let us first demonstrate how one can
obtain
the CFT results of section ~\ref{sec:CFT}
using only the effective field theory.
Masses $m_{ab}$ have been calculated in, for example, Refs.~\cite{Ghilencea:2002da,Ghilencea:2002by},
but we shall extend this approach to the computation of kinetic mixing $\chi_{ab}$.
To do this, we consider the action of the brane and the supergravity
fields \cite{Cheung:1997az,Minasian:1997mm,Bachas:1999um,Polchinski:1998rr}:
\begin{eqnarray}
\label{DBIact}
S_{\rm DBI}&=& \mu_p \int d^{p+1} x e^{-\Phi} \sqrt{-\det g+ 2\pi \ap F + B}  \\
&\approx &
\int d^{p+1} x \mu_p e^{-\Phi} \sqrt{-g}
-\frac{1}{4} \mu_p e^{-\Phi}  \sqrt{-g} \bigg((2\pi \ap)^2 F_{\mu \nu} F^{\mu \nu}
+ 2(2\pi \ap) F_{\mu \nu} B^{\mu \nu} + B_{\mu \nu} B^{\mu \nu} \bigg), \nonumber \\
\label{WZact}
S_{\rm WZ} &=& \mu_p \int_{Dp} \sum_{q} C_q \wedge tr \exp (2\pi \ap F + B) \wedge
\sqrt{\frac{\hat{A} (4\pi^2 \ap R_T)}{\hat{A} (4\pi^2 \ap R_N)}},  \\
S_{\rm R} &=&  - \frac{1}{4\kappa_{10}^2} \int d^{10} x (-\det G)^{1/2} \bigg( |F_1|^2 + |\tilde{F}_3|^2
+ \frac{1}{2} |\tilde{F}_5|^2 \bigg ), \\
S_{\rm NS} &=& - \frac{1}{4\kappa_{10}^2} \int d^{10} x (-\det G)^{1/2}  e^{-2\Phi} |H_3|^2 \, ,
\label{ActionList}\end{eqnarray}
where $A_\mu$ is a gauge field, $C_q$ are the R-R forms and $B_2$ is the NS-NS 2-form and the
field-strengths are defined as
\begin{eqnarray}
F&=& d A, \nonumber \\
F_{q+1} &=& d C_q, \nonumber \\
H_3 &=& d B_2, \nonumber \\
\tilde{F}_3 &=& F_3 - C_0 \wedge H_3, \nonumber \\
\tilde{F}_5 &=& F_5 - \frac{1}{2} C_2 \wedge H_3 + \frac{1}{2} B_2 \wedge F_3, \nonumber \\
*_{10} \tilde{F}_5 &=& \tilde{F}_5.
\end{eqnarray}
Note that the 2-form $F$ in the first equation above is the usual $F_{\mu\nu}$ and should not be
confused with the R-R field strength $F_2=d C_1$ in the second equation.
Also $\mu_p = \sqrt{2\pi} (4\pi^2 \ap)^{-\frac{1+p}{2}}$ is the brane tension,
and $2\kappa_{10}^2 =  (\ap)^4 (2\pi)^7$.

Already from the Dirac-Born-Infield (DBI) action \eqref{DBIact} it is clear that the $B$-field
can mediate kinetic mixing. In addition, for D$p$-branes, a $p-1$-form $C_{p-1}$ couples
in \eqref{WZact} to the gauge fields, which can mediate between branes of the same dimensionality.

\subsection{A simple case without fluxes}

The CFT results for kinetic mixing, e.g., Eq.~\eqref{D3D3barmixing} in the D3-\dbar{3} system
on a toroidal background, are applicable for backgrounds without flux vacuum expectation values.
We will now calculate
the same results using the supergravity approach based on the DBI action \eqref{DBIact}.

The vertices for the antisymmetric tensor $B_{\mu \nu}$ and $A_{\rho}$ are
\begin{equation}
 \frac{1}{2} 2\pi \ap \mu_p g_s^{-1} (k_{\mu} g_{\nu\rho} - k_{\nu} g_{\mu\rho}) \delta(\Sigma_{p}),
\end{equation}
for a $p$-brane of worldvolume $\Sigma_p$.
The propagator for a component of $B_{\mu \nu}$, $\mu, \nu \in \{0,1,2,3\}$ is straightforward to write down:
the diagonal part of the propagator is
\beq
G_{\mu\nu;\rho\sigma} (k_4,y_0,y_1) =  \delta_{\mu \rho} \delta_{\nu \sigma}
\frac{2g_s^2 \kappa_{10}^2}{V_6} \sum_{k_6}
\frac{\exp \left[ \imi k_6 \cdot (y_1 - y_0) \right]}{|k_4|^2 + |k_6|^2}.
\eeq
Here $k_4$ and $k_6$ are the 4-dimensional and the transverse 6-dimensional momenta
(w.r.t $\mu, \nu \in \{0,1,2,3\}$
of $B_{\mu \nu}$),
and $y_1-y_0$ is the 6-dimensional distance vector in the transverse space.

The $B$-field induced contribution to the 2-point function of gauge fields is
\begin{multline}
\bra A^{a_1}_{\mu_1} A^{b_1}_{\nu_1} \ket_B = \,
\frac{\delta}{\delta A^{a_1}_{\mu_1}}\frac{\delta}{\delta A^{b_1}_{\nu_1}}\,
\tr_1 \lambda_a \tr_2 \lambda_b \frac{1}{2} \frac{1}{\ap} \frac{(2\pi \ap)^3}{V_6}
(4\pi^2 \ap)^{\frac{3-p_a}{2}}(4\pi^2 \ap)^{\frac{3-p_b}{2}} \bigg[ A^{a}_{\mu} A_{b}^{\mu}
V_{Dp_a} V_{Dp_b} \\
+  (k^2_{4} A^{a}_{\mu} A_{b}^{\mu} - k_{4} \cdot A^a k_{4} \cdot A^b)
\int d^{p_a-3} y d^{p_b-3} y_1 \sum_{k_6}
\frac{\exp \left[ \imi k_6 \cdot (y_b - y_a) \right]}{ |k_6|^2} \bigg] .
\end{multline}
This is the contribution from the $B$-field only. On the torus, there will also be contributions from
$C_{p-1}$-forms but only if $p_a = p_b$.
In this case, for rectangular tori, one can show that for brane-brane mixing the $C$-form contribution
is equal and opposite to the $B$-contribution, exactly cancelling it;
while for brane-antibrane mixing they are equal in sign and magnitude, simply multiplying the above by two.
To reproduce the results of Ref.~\cite{Abel:2003ue},
consider brane-antibrane mixing on untwisted tori, so that in Neumann-Dirichlet directions the integrals
in the above become delta functions; for $p \ne q$ we obtain
\beq
\chi_{ab} = g_a g_b \tr_1 \lambda_a \tr_2 \lambda_b \frac{1}{2\pi}\frac{l_s^6}{V_6}
\frac{(V_a V_b)}{l_s^{p_a + p_b - 6}}\sum_{n_i}
\frac{\prod_{i=1}^{N_{\rm DD}} \exp \left[  2 \pi \imi
\frac{n_i}{R_i}  (y_b^i - y_a^i) \right]}{ \sum_{i=1}^{N_{\rm DD}} n_i^2 l_s^2/R_i^2},
\eeq
where $l_s^2 = 2\pi \ap$, and $N_{\rm DD}$ is the number of Dirichlet-Dirichlet directions.
For $p=q$ and parallel
brane and antibrane the final result is twice the above formula.
This agrees with the results of \cite{Abel:2003ue} and for $p=3=q$ with our earlier CFT-derived
Eq.~\eqref{D3D3barmixing} (for the D3-\dbar{3} system
in the present context of an untwisted toroidal background).

For more general models, where the compact manifold is not a rectangular torus, we consider the action for
a single component of $B_{\mu \nu}$ which we shall denote $\phi$, and neglect the transverse modes
(the calculation for the $C$-field being identical apart from a minor modification to
the vertices):
\beq
S = \frac{1}{2\kappa_{10}^2}\int  \frac{d^4 x}{(2\pi)^4} \int_{M^6} e^{-2\Phi}\bigg( \frac{1}{2}
k_4^2 \phi^2 + \frac{1}{2} d^{(6)} \phi \wedge \star_6 d^{(6)} \phi \bigg).
\eeq
For a constant dilaton $e^{-2\Phi} = g_s^{-2}$, the Green functions therefore obey
\beq
(k^2_4 + \Delta_6  ) G_{\mu \nu; \rho \sigma} (y_0,y_1) =
\delta_{\mu \rho} \delta_{\nu \sigma} 2\kappa_{10}^2 g_s^2 \delta (y_1 - y_0),
\eeq
where $\Delta_6 = d \star_6 d + (d d \star_6) $ is the Laplacian on the compact manifold.
To solve the above, we may expand in terms of orthogonal, normalised, eigenfunctions $\phi_n$ of the
Laplacian (a Hermitian operator provided the manifold admits a Hermitian metric) with eigenvalue $\alpha_n$,
in terms of which the Green function is
\beq
G_{\mu \nu; \rho \sigma} = \delta_{\mu \rho} \delta_{\nu \sigma} 2\kappa_{10}^2 g_s^2 \sum_{n}
\frac{b_n (y_1) \phi_n^{*} (y_1)}{\alpha_n + k_4^2} \phi_n (y_0),
\eeq
where $b_n (y_1)$ is the weight function. Clearly there is a contribution to the mass term only
when $\alpha_n = 0$, and in all the other contributions, to calculate
the mixing we may set $k_4^2 = 0$ in the denominator.

Since we are only interested in the fields with indices in the noncompact dimensions
(that will couple to the gauge field) the Green function is treated as a zero-form on the compact space.
This means that the zero modes of the Laplacian, being harmonic forms, are in one to one correspondence
with $H^0 (M, \mathbb{R})$:
for Calabi-Yau manifolds the only zero mode is the constant solution $\phi_0$.
This then implies that the result for the mass mixing in Eq.~(\ref{m2}) applies quite generally and is not particular to the torus.

As a simple example, we apply this to the case of a product of tilted tori,
where the metric is
$ds^2 = \sum_{\kappa=1}^3 (2\pi)^2 \frac{T_2^{\kappa}}{U_2^{\kappa}} dz^{\kappa} d\bar{z^{\kappa}}$
and the periodicities are $z^{\kappa} \sim z^{\kappa}+1, z^{\kappa} \sim z^{\kappa}+U^{\kappa}$. We then have
the Laplacian
\beq
\Delta_6 = \partial_i (\sqrt{g_6} g^{ij}_6 \partial_j) =
\sqrt{g_6} \frac{U_2^{\kappa}}{(2\pi)^2 T_2^{\kappa}} \partial_{\kappa} \bar{\partial}_{\kappa} .
\eeq
The eigenfunctions are easily found to be $\frac{1}{\sqrt{V_6}} \prod_{\kappa} e^{2\pi
\imi p^{\kappa} x^{\kappa}} e^{2 \pi \imi q^{\kappa} y^{\kappa}}$,
where $z^{\kappa} = x^{\kappa} + U^{\kappa} y^{\kappa}$ and
$x^{\kappa} \sim x^{\kappa} + 1, y^{\kappa} \sim y^{\kappa} + 1$;
the weight function is just $\sqrt{g_6}$ so we have
\beq
G_{\mu \nu; \rho \sigma } = \delta_{\mu \rho} \delta_{\nu \sigma}\frac{2\kappa_{10}^2 g_s^2}{V_6}
\sum_{p^{\kappa},q^{\kappa}}
\frac{\prod_{\kappa} e^{\frac{2\pi \imi}{U_2} {\rm Im}( p^{\kappa} (y^{\kappa}_0 - y_1) \ov{U}^{\kappa}
+ q^{\kappa} (y^{\kappa}_0 - y_1) )} } {\sum_{\kappa} \frac{1}{T_2^{\kappa} U_2^{\kappa}} |q^{\kappa}
+ p^{\kappa} \ov{U}^{\kappa} |^2 }.
\eeq
Using the same vertices as before, we clearly obtain the same result as the string calculation
of kinetic mixing in the ${\rm D}3$-$\overline{\rm D}3$ system
found in Eq.~(\ref{D3D3barmixing}).

\subsection{Inclusion of vacuum expectation values for fluxes}
\label{VFlux}

In type IIB model building it is usually required to include vacuum expectation values (vevs)
for the three-form fluxes in order to stabilise the moduli, and so in order to go further,
we need to incorporate this into our calculation of kinetic mixing.

Let us begin with a simple observation. The effect of the fluxes in some of the most interesting cases,
including
KKLT models \cite{KKLT,Grana:2005jc}, is that the metric is warped in the vicinity of the standard model branes:
\beq
ds^2 = e^{2A(y)} \eta_{\mu \nu} dx^{\mu} dx^{\nu} + e^{-2A(y)} g_{mn} dy^m dy^n .
\eeq
If we now restrict our attention to (anti) D3-branes then we can immediately see that,
since the coupling of the gauge fields to the antisymmetric tensor and the R-R two-form is
classically conformal, the
kinetic mixing cannot depend upon the warp factor. Hence all of the modification to the previous
cases will derive from the Green function.

To proceed, we split the fields into $B_2 = B_2^{(4)} + B_2^{(6)} + B_2^{(46)}$
(and similarly for the two-form $C_2$) where the superscripts are as follows:
(4) indicates both are space-time indices, (6) indicates both are internal, (46) indicates one of each.
As we have mentioned, only the components $B_2^{(4)}, C_2^{(4)}$ can mediate the mixing, but now we
wish to give a vev to the components $B_2^{(6)},C_2^{(6)}$. There may also be a vev for the five-form
field $F_5$, but this does not contribute -- on the other hand, the contribution of the two-form vevs
to $\tilde{F}_5$ are crucial.

Importantly, the vevs for $B_2^{(6)},C_2^{(6)}$ generate masses for the two-form fields $B_2^{(4)}, C_2^{(4)}$.
In fact, from Eqs.~\eqref{DBIact}-\eqref{ActionList} we can read off the kinetic and mass terms for $B_2$,
\begin{multline}
S = \frac{1}{2\kappa_{10}^2} \int d^4 x d^6 y \bigg( | \tau |^2 + |C_2^{(6)}|^2 \bigg)
\frac{1}{2} |H_3^{(4)} |^2 + \frac{1}{8}|B_2^{(4)}|^2 |F_3^{(6)}|^2 \\
+ | \tau |^2 | d^{(6)} B_2^{(4)}|^2 + \frac{1}{8}| C_2^{(6)} \wedge d^{(6)} B_2^{(4)}|^2 \, ,
\label{closedlag}\end{multline}
while for $C_2$ we have similarly
\begin{multline}
S = \frac{1}{4\kappa_{10}^2} \int d^4 x d^6 y \bigg( 1 + \frac{1}{8}|B_2^{(6)}|^2 \bigg)  |F_3^{(4)} |^2
+ \frac{1}{8}|C_2^{(4)}|^2 |H_3^{(6)}|^2 \\
+ | d^{(6)} C_2^{(4)}|^2 + \frac{1}{8}| B_2^{(6)} \wedge d^{(6)} C_2^{(4)}|^2 \, .
\label{CLag}\end{multline}
Thus fluxes generate masses for the two-form fields; from the string point of view we have stabilised
the moduli.

Since the $B_2^{(4)}, C_2^{(4)}$ fields are now massive this has an
effect on the mixing of the $U(1)$ factors which these fields
mediate.

Ignoring the non-compact dimensions' kinetic terms, for a component $\phi$ of  $C_{\mu \nu}$ we have
\beq
\mathcal{L} = \frac{e^{-2A}\sqrt{g} }{2\kappa_{10}^2}\bigg[ g^{m n} \partial_m \phi \partial_n \phi
+ \frac{1}{8}| B_2^{(6)} \wedge d^{(6)}\phi|^2 + \frac{1}{8}|H_3^{(6)}|^2 \phi^2 \bigg],
\label{scalareq}
\eeq
where the last term is a mass of $\phi$. We can estimate the magnitude of the effective
$\phi$-mass by considering that $H_3$ and $F_3$ are defined as fluxes threading
three-cycles \cite{Grana:2005jc},
\begin{eqnarray}
\frac{1}{(2\pi)^2 \ap} \int_{A_K} H_3 &=& m^K, \nonumber \\
\frac{1}{(2\pi)^2 \ap} \int_{B^K} F_3 &=& e_K,
\end{eqnarray}
where $m^K, e_K$ are integers and $K=1..h^3$. Thus we can estimate that
\beq
H_3, F_3 \sim n l_s^2 /V_3,
\label{RoughHFMass}\eeq
for some integer $n$ and different three-cycle volumes $V_3$.
Provided that the cycles threaded by the flux are larger than the string scale, we expect the second
term in \eqref{scalareq} to be less significant, and $\phi$ should behave like a massive scalar
with a characteristic length given by $L \sim V_3/(nl_s^2)$, i.e. the Green functions for the
two-form fields behave as
\beq
G_{\mu \nu;\rho\sigma} (y) \propto \delta_{\mu \rho}\delta_{\nu \sigma} e^{-y n l_s^2/V_3}\, .
\label{WaveHands}\eeq
The resulting interaction is a ``Yukawa type'' interaction (whose exponential form derives from the mass of the
mediating scalar and has nothing to do with the warping). Thus we expect to be able to probe much of the
compact manifold; for $V_3$ of ${\cal O}(100)$ we can probe ${\cal O}(1000)$ string length distances.
Note that the three-cycles that are threaded will usually be different for $H_3$ and $F_3$. If one of
them is much smaller (or the fluxes larger)
as is usually the case, then
the corresponding two-form is subdominant in the generation of kinetic mixing.

With the full dependence on fluxes to hand, we now proceed to consider specific warped models
where we can solve the equations exactly, in order to verify the general behaviour anticipated above.
We will consider two cases.
The first in the following section
is a simplified Randall-Sundrum model which demonstrates that the kinetic mixing is
independent of the warping but depends only on the induced mass via the Green function.
The second model is the more ``realistic'' case of kinetic mixing in the Klebanov-Tseytlin throat.

\section{Randall-Sundrum models}
\label{sec:RS}

Randall-Sundrum (RS) models \cite{Randall:1999vf,Randall:1999ee} involve branes embedded in a slice of
$AdS_5$. They may be considered
as dimensionally reduced string models, or as legitimate phenomenological models in their own right.
Since they involve a warped hidden dimension, they are candidates for use as a toy for examining
kinetic mixing in a non-trivial background, but still with hope of tractability. Some related work
has been performed in \cite{Gherghetta:2000qt,Gherghetta:2000kr,Maity:2007un}, under the assumption
that the matter fields are not confined to the branes but have wavefunctions extending throughout the
fifth dimension. In their case, there were no additional fields, since the wavefunction overlaps
contributed to kinetic mixing.
We shall rather consider a more string-inspired scenario, where matter fields are confined to branes,
and thus shall introduce a string-inspired $B$-field.
The metric for the model is taken to be
\beq
ds^2 = e^{-2k|y|} \eta_{\mu \nu} d x^\mu d x^\nu + d y^2,
\eeq
with $k$ a parameter of the order of the Planck scale.

We shall consider the standard model brane to be a D3-brane at a position $y = 0$ in the hidden dimension,
and a hidden brane at some position $y_1 = \pi R$. We shall suppose that there is a massless $U(1)$
field supported upon each, and shall then calculate the mixing. (More generally, we could consider
several branes with the $U(1)$ split between them so as to make non-anomalous combinations; the mixing
would then be given by calculating the mutual differences.) The Lagrangian of our model is taken to be
\begin{eqnarray}
\mathcal{L} &=& \mathcal{L}_{\rm bulk} + \mathcal{L}_{\dthree} +  \mathcal{L}_{\dthreebar}, \nonumber \\
\mathcal{L}_{\rm bulk} &=& \frac{M_5^3}{2 g^4} \int \frac{-1}{2 } \d B \wedge *_5 \d B
+ \frac{1}{2} m^2 B \wedge *_5 B, \nonumber \\
\mathcal{L}_{\dthree} &=& \frac{1}{4g^2} \int_{\mathrm{ D3}} \frac{1}{2\pi \ap} F \wedge *_4 B
+ \frac{1}{(2\pi\ap)^2} B \wedge *_4 B, \nonumber \\
 \mathcal{L}_{\dthreebar} &=& - \mathcal{L}_{\dthree} .
\end{eqnarray}
Here we have introduced a $B$-field which will mediate the mixing. The coupling of the $B$-field
to the gauge field is specified by the Dirac-Born-Infeld action, but we should point out that it is
necessary to introduce three parameters into the model: the coupling of the kinetic term $M_5$, the
mass-like parameter $m$ and the string mass. If we imagine the above to be derived from a string model,
then we expect $M_5$ to be related to Planck's constant and the volume of the compactification, and $m$
to be determined by the fluxes; the string scale however generally exists as a free parameter to be
determined by experiment. It is tempting to relate the $M_5$ coupling to the parameters already extant
in the RS scenario; we shall examine the consequences of this later.

To calculate the mixing, as already mentioned we require the Green function, and thus we derive the
(very simple) equations of motion
\beq
\bigg[ e^{2 k |y|} \eta^{\alpha \beta} \partial_\alpha \partial_\beta + \partial_5 \partial_5
- m^2 \bigg] B_{\mu \nu}^{(4)} = 0 \ .
\eeq
However, from the above action we also find boundary conditions for the $B$-field at the brane:
\beq
\partial_y B_{\mu \nu}^{(4)} -\frac{M_s^4}{M_5^3} B_{\mu \nu}^{(4)} |_{y=0,\pi R} = 0 \, .
\eeq
Since we shall effectively be finding a one-dimensional Green function, these conditions become important;
if we were considering a higher dimensional model we could neglect them and consider only the periodic
boundary conditions of the compact space.

\subsection{Green functions in one dimension}

The Green functions are straightforward to find for the above action, and the result is a propagator
that actually dies more rapidly than the equivalent RS solution at large distances. The procedure for
computing them is adapted from Ref.~\cite{Gherghetta:2000kr}\footnote{In fact, they derived the propagator
for the Randall-Sundrum model at general positions in the bulk at finite momenta. It is possible to
extract the information we need from their equation (62) in Ref.~\cite{Gherghetta:2000kr} by setting the
momentum and the parameter $s$ to zero. However, it is actually easier and more transparent to
rederive the expression we need.} as follows. Let
\beq
\Delta G(y,y') = \delta(y-y')
\eeq
define the Green function (note the loss of translational invariance due to the positions of the branes).
Now decompose it into ``advanced'' and ``retarded'' components:
\beq
G (y,y') = \theta(y-y') G_{>}(y,y') + \theta(y'-y) G_{<} (y,y')\, ,
\eeq
where $G_{>},G_{<}$ satisfy the homogeneous equation, and we must impose matching conditions at $y=y'$.
Writing this as
\beq
\partial_y (f(y) \partial_y G(y,y')) - h(y) G(y,y') = k(y) \delta(y-y')
\eeq
(with the redundancy deliberate) we obtain the continuity condition $G_{>} (y,y) = G_{<} (y,y)$, and
\beq
\partial_y G_{>} (y,y) - \partial_y G_{<} (y,y) = \frac{k}{2f},
\eeq
which sets the normalisation of the propagator. We now solve by separation of variables:
\begin{eqnarray}
G_{<} (y,y') &\equiv& A_{<} (y') \tilde{G}_{<} (y), \nonumber \\
G_{>} (y,y') &\equiv& A_{>} (y') \tilde{G}_{>} (y),
\end{eqnarray}
to find the equations
\begin{eqnarray}
A_{<} (y) \tilde{G}_{<}^{\prime} (y) - A_{>} (y) \tilde{G}_{>}^{\prime} (y) &=&\frac{k}{2f}, \nonumber \\
A_{<} (y) \tilde{G}_{<} (y) - A_{>} (y) \tilde{G}_{>} (y) &=& 0.
\end{eqnarray}
These allow us to find the $A$ functions. Now, what we desire is $G(y_0,y_1)$, where $y_0,y_1$ are
the coordinates of the branes. This is given by $A_{<} (y_1) \tilde{G}_{<} (y_0)$, but these contain
values at the boundaries: our boundary conditions are
\begin{eqnarray}
\partial \tilde{G}_{<} (y_0) &=& r \tilde{G}_{<} (y_0), \nonumber \\
\partial \tilde{G}_{>} (y_1) &=& s \tilde{G}_{>} (y_1),
\end{eqnarray}
and we then find the result
\beq
G(y_0,y_1) = \frac{k}{2f}(y_1) \frac{\tilde{G}_{<} (y_0)}{\tilde{G}_{<}^{\prime} (y_1)
- s \tilde{G}_{<} (y_1)}\, .
\eeq
This is particularly simple to solve numerically; we simply solve one homogeneous equation for
$\tilde{G}_{<} (y)$ with initial conditions
\begin{eqnarray}
\tilde{G}_{<} (y_0) &=& C, \nonumber \\
\tilde{G}_{<}^{\prime} (y_0) &=& r C,
\end{eqnarray}
and the result is independent of the choice of $C$.

For the massive RS action, we can actually solve exactly to obtain
\beq
G(y_0,y_1) = \frac{4g^2 }{M_5^3 m} \frac{1}{\sinh m \pi R} \frac{1}{\left(
1- \frac{M_s^8}{ M_5^6 m^2}\right) } .
\eeq
This gives mixing
\beq
\chi = g_a g_b \frac{32 M_s^4}{M_5^3 m} \frac{1}{\sinh m \pi R} \frac{1}{\left(
1- \frac{M_s^8}{ M_5^6 m^2}\right)} .
\eeq
It is tempting to identify $M_5$ with $M$ from the existing RS parameters,
where $e^{-k\pi R} = M_{\rm SUSY}/M_{\rm Pl}$, (so $\pi k R \sim 16 \log 10 \sim 37 $),
$M_{\rm Pl}^2 \approx M^3/k$.
We also make the assumption that $M_s^8/(M_5^6 m^2)$ is small. For concreteness, we will take an
intermediate string mass of $M_s=\sqrt{M_{\rm SUSY}M_{\rm Pl}}$.
This gives
\begin{eqnarray}
\chi &\approx& g_a g_b \frac{32}{\log \frac{M_{\rm Pl}}{M_{\rm SUSY}}} \frac{M_s^4}{M_{\rm Pl}^2}
\frac{\pi R}{m}
\frac{1}{\sinh \pi m R} \nonumber \\
&\approx& g_a g_b \frac{32}{37}   \times \frac{M^2_{\rm SUSY} \pi R}{m} \frac{1}{\sinh \pi m R}\, .
\end{eqnarray}
In the limit that $mR \ll 1$, we have
\beq
\chi  \sim g_a g_b  \times \frac{M^2_{\rm SUSY} }{m^2} \, .
\eeq
For gauge couplings of order unity, we see
that values of $m \sim 10^{4} M_{\rm SUSY}$ leads to a mixing that is observable in the near future.
Comparing this to equation (\ref{RoughHFMass}), if the flux mass is related to a three-cycle in
some compact space we should have typical length for a wrapping cycle of ${\cal O}(10^2 l_s)$.

In the opposite limit, $mR \gg 1$, one gets the expected exponential suppression due to the
non-zero mass:
\beq
\chi  \sim g_a g_b  \times \frac{M^2_{\rm SUSY}}{m^2} \, (m \pi R)\,  e^{-m \pi R} \, .
\eeq
Thus in RS backgrounds the kinetic mixing can reasonably take any value between zero and
the experimental limits,
depending on the configuration.

\section{Kinetic mixing on the Klebanov-Tseytlin throat}
\label{sec:KT}

We now turn to an example of a Calabi-Yau manifold for which the metric is known, the Klebanov-Tseytlin
throat \cite{Klebanov:2000nc,Herzog:2001xk}. This is a model of a warped throat region, as found in
KKLT models \cite{KKLT}. In this model, there are flux vacuum expectation values, but the back reaction
of the flux upon the metric is not fully included. It can thus be seen as an approximation to the
Klebanov-Strassler solution \cite{Klebanov:2000hb}, where we introduce by hand an ``infrared''
cutoff $r_s$ to model the effect of removing the conical singularity; there is also an ``ultraviolet''
cutoff $r_0$ in both models to render the solution compact (so that $r_s < r <r_0$).
In the near-horizon limit it reduces to
an RS model, and so we might expect a similar exponential damping effect to occur here. Let us now see
if this is the case.

Consider the metric on a general cone:
\beq
ds^2 = h^{-1/2} (r) \eta_{\mu \nu} dx^{\mu} dx^{\nu} + h^{1/2}(r) (dr^2 + r^2 ds_{M}^2)
\eeq
where $M$ is Sasaki-Einstein maifold. If we put a cutoff at some large radius $r_0$ we can consider the above to
be part of a larger compact Calabi-Yau manifold (henceforth the bulk). If we wish to use this as a particle
physics model, we must then include fluxes to stabilise the various complex structure moduli, which will
warp the throat. Moreover, we must consider how to embed the standard model on branes: we may have either
D3-branes at the singularity \cite{Cascales:2003wn} (where we must generalise the above, with for example
an orbifold projection) or D3/D7 branes wrapping appropriate cycles elsewhere, either in the throat or the bulk,
with {\dbar}3-branes at the tip \cite{Balasubramanian:2005zx,Conlon:2005ki,Conlon:2006wz}
(which can uplift an AdS vacuum to a dS one). Consistency of the model also requires orientifold
planes, but they may be present either at the tip of the throat or in the bulk, in the latter case
necessitating an image throat.

The metric on the Klebanov-Tseytlin solution is (conventionally written in the Einstein frame, defined by $ds^{2}_{\rm Einstein} = \sqrt{g_s} ds^{2}  $)
\begin{eqnarray}
h(r) &=& \frac{81(g_s M \ap)^2 \log r/r_s}{8r^4} \nonumber \\
&=& \frac{27 (\ap)^2 (2 g_s N + 3(g_s M)^2 \log(r/r_0) + 3 (g_s M)^2 /4)}{8r^4},
\nonumber \end{eqnarray}
where $M$ is the number of fractional D5 branes wrapped on a compact $S^3 \subset T^{1,1}$
at the tip of the throat, and
\beq
ds_M^2 = ds_{T^{1,1}}^2 = \frac{1}{9} \left( d \psi + \sum_{i=1}^2 \cos \theta_i d\phi_i \right)^2
+ \frac{1}{6} \sum_{i=1}^2 \left(d\theta_i^2 + \sin \theta_i d\phi_i^2 \right) \, .
\eeq
To calculate the kinetic mixing, we consider the dynamics of the R-R
two-form\footnote{The dynamics of the $B$-field is complicated by its coupling to $C^{(6)}_2$,
which admits no globally smooth vev due to the non-compact nature of the conifold. If desired,
the same analysis can be performed in the large distance limit, giving a similar,
but crucially not identical, result.}, $C^{(4)}_2$.
We start from the action (\ref{CLag}),
and make the assumption that in the throat region, the radial Kaluza-Klein modes are much
less significant that the longitudinal modes; this is reasonable since the throat will be long and thin.
Again writing $\phi$ for a component of
$C_{\mu \nu}$, the action for the relevant term then reduces to
\beq
S = \frac{1}{4g_{s}\kappa_{10}^2} \int d^4 x \partial_r \phi \partial_r \phi
(|dr|^2 + \frac{1}{8}| B_2^{(6)} \wedge dr|^2) + \frac{1}{8} \phi^2 |H_3^{(6)}|^2 \, .
\eeq
The vacuum expectation values for the fluxes are taken to be
\begin{eqnarray}
F_3 &=& \frac{M\ap}{2} \omega_3, \nonumber\\
H_3 &=& \frac{3g_s M\ap}{2r} dr \wedge \omega_2, \nonumber \\
B_2 &=& \frac{3g_s M\ap}{2} \log (r/r_0) \ \omega_2,
\end{eqnarray}
where
\begin{eqnarray}
\omega_3 &=& (d\psi + \cos \theta_1 d\phi_1 + \cos \theta_2 d \phi_2 )\wedge \omega_2, \nonumber \\
\omega_2 &=& \frac{1}{2} (\sin \theta_1 d\theta_1 \wedge d\phi_1 + \sin \theta_2 d\theta_2 \wedge d \phi_2),
 \nonumber \\
\omega_2 \wedge \omega_3 &=& 54 {\rm Vol}(T^{1,1}), \nonumber \\
\omega_3 &=& -\frac{3}{r} *_6 (dr \wedge \omega_2) \, .
\end{eqnarray}
The action then becomes
\beq
S = \frac{3\pi^3 (M\ap )^2g_s}{4\kappa_{10}^2} \int d^4 x dr \,\,
\partial_r \phi \partial_r \phi ( 2 r \log r/r_s  + r (\log r/r_0)^2 ) +  \phi^2 \frac{1}{r} \, .
\eeq
Using the variable $y = \log r/r_s$, we then have
\beq
S = \frac{3\pi^3 (M\ap)^2 g_s}{4\kappa_{10}^2} \int d^4 x dy \,\,
\partial_y \phi \partial_y \phi ( 2 y + (y -y_0)^2) + \phi^2.
\eeq
Note that in the small distance limit, this becomes the equation for a Randall-Sundrum model
with a constant mass $\phi$-field, with
\[\frac{M^3_5}{g^2} = \frac{3\pi^3 (M\ap)^2g_s}{\kappa_{10}^2} y_0^{-2}\, ,\] and $m^2 = y_0^{-2}$
(note that the dimensions are different). This is quite indicative: in the Klebanov-Tseytlin throat
the effective mass is smaller than $1$, and moreover for branes in the throat the separation will
never exceed the inverse mass -- so we do not expect a large suppression.

\begin{figure}
\begin{center}
\setlength{\unitlength}{0.240900pt}
\ifx\plotpoint\undefined\newsavebox{\plotpoint}\fi
\sbox{\plotpoint}{\rule[-0.200pt]{0.400pt}{0.400pt}}%
\begin{picture}(1500,900)(0,0)
\font\gnuplot=cmr10 at 10pt
\gnuplot
\sbox{\plotpoint}{\rule[-0.200pt]{0.400pt}{0.400pt}}%
\put(181.0,123.0){\rule[-0.200pt]{4.818pt}{0.400pt}}
\put(161,123){\makebox(0,0)[r]{ 0}}
\put(1419.0,123.0){\rule[-0.200pt]{4.818pt}{0.400pt}}
\put(181.0,228.0){\rule[-0.200pt]{4.818pt}{0.400pt}}
\put(161,228){\makebox(0,0)[r]{ 0.5}}
\put(1419.0,228.0){\rule[-0.200pt]{4.818pt}{0.400pt}}
\put(181.0,334.0){\rule[-0.200pt]{4.818pt}{0.400pt}}
\put(161,334){\makebox(0,0)[r]{ 1}}
\put(1419.0,334.0){\rule[-0.200pt]{4.818pt}{0.400pt}}
\put(181.0,439.0){\rule[-0.200pt]{4.818pt}{0.400pt}}
\put(161,439){\makebox(0,0)[r]{ 1.5}}
\put(1419.0,439.0){\rule[-0.200pt]{4.818pt}{0.400pt}}
\put(181.0,544.0){\rule[-0.200pt]{4.818pt}{0.400pt}}
\put(161,544){\makebox(0,0)[r]{ 2}}
\put(1419.0,544.0){\rule[-0.200pt]{4.818pt}{0.400pt}}
\put(181.0,649.0){\rule[-0.200pt]{4.818pt}{0.400pt}}
\put(161,649){\makebox(0,0)[r]{ 2.5}}
\put(1419.0,649.0){\rule[-0.200pt]{4.818pt}{0.400pt}}
\put(181.0,755.0){\rule[-0.200pt]{4.818pt}{0.400pt}}
\put(161,755){\makebox(0,0)[r]{ 3}}
\put(1419.0,755.0){\rule[-0.200pt]{4.818pt}{0.400pt}}
\put(181.0,860.0){\rule[-0.200pt]{4.818pt}{0.400pt}}
\put(161,860){\makebox(0,0)[r]{ 3.5}}
\put(1419.0,860.0){\rule[-0.200pt]{4.818pt}{0.400pt}}
\put(181.0,123.0){\rule[-0.200pt]{0.400pt}{4.818pt}}
\put(181,82){\makebox(0,0){ 0}}
\put(181.0,840.0){\rule[-0.200pt]{0.400pt}{4.818pt}}
\put(338.0,123.0){\rule[-0.200pt]{0.400pt}{4.818pt}}
\put(338,82){\makebox(0,0){ 2}}
\put(338.0,840.0){\rule[-0.200pt]{0.400pt}{4.818pt}}
\put(496.0,123.0){\rule[-0.200pt]{0.400pt}{4.818pt}}
\put(496,82){\makebox(0,0){ 4}}
\put(496.0,840.0){\rule[-0.200pt]{0.400pt}{4.818pt}}
\put(653.0,123.0){\rule[-0.200pt]{0.400pt}{4.818pt}}
\put(653,82){\makebox(0,0){ 6}}
\put(653.0,840.0){\rule[-0.200pt]{0.400pt}{4.818pt}}
\put(810.0,123.0){\rule[-0.200pt]{0.400pt}{4.818pt}}
\put(810,82){\makebox(0,0){ 8}}
\put(810.0,840.0){\rule[-0.200pt]{0.400pt}{4.818pt}}
\put(967.0,123.0){\rule[-0.200pt]{0.400pt}{4.818pt}}
\put(967,82){\makebox(0,0){ 10}}
\put(967.0,840.0){\rule[-0.200pt]{0.400pt}{4.818pt}}
\put(1125.0,123.0){\rule[-0.200pt]{0.400pt}{4.818pt}}
\put(1125,82){\makebox(0,0){ 12}}
\put(1125.0,840.0){\rule[-0.200pt]{0.400pt}{4.818pt}}
\put(1282.0,123.0){\rule[-0.200pt]{0.400pt}{4.818pt}}
\put(1282,82){\makebox(0,0){ 14}}
\put(1282.0,840.0){\rule[-0.200pt]{0.400pt}{4.818pt}}
\put(1439.0,123.0){\rule[-0.200pt]{0.400pt}{4.818pt}}
\put(1439,82){\makebox(0,0){ 16}}
\put(1439.0,840.0){\rule[-0.200pt]{0.400pt}{4.818pt}}
\put(181.0,123.0){\rule[-0.200pt]{0.400pt}{177.543pt}}
\put(181.0,123.0){\rule[-0.200pt]{303.052pt}{0.400pt}}
\put(1439.0,123.0){\rule[-0.200pt]{0.400pt}{177.543pt}}
\put(181.0,860.0){\rule[-0.200pt]{303.052pt}{0.400pt}}
\put(40,491){\makebox(0,0){$\frac{3 M^2}{32g_{a}g_{b}}\chi$}}
\put(810,21){\makebox(0,0){Brane Displacement}}
\put(194,781){\usebox{\plotpoint}}
\multiput(194.58,735.06)(0.492,-14.153){21}{\rule{0.119pt}{11.067pt}}
\multiput(193.17,758.03)(12.000,-306.031){2}{\rule{0.400pt}{5.533pt}}
\multiput(206.58,437.54)(0.493,-4.343){23}{\rule{0.119pt}{3.485pt}}
\multiput(205.17,444.77)(13.000,-102.768){2}{\rule{0.400pt}{1.742pt}}
\multiput(219.58,333.97)(0.492,-2.349){21}{\rule{0.119pt}{1.933pt}}
\multiput(218.17,337.99)(12.000,-50.987){2}{\rule{0.400pt}{0.967pt}}
\multiput(231.58,282.50)(0.493,-1.250){23}{\rule{0.119pt}{1.085pt}}
\multiput(230.17,284.75)(13.000,-29.749){2}{\rule{0.400pt}{0.542pt}}
\multiput(244.58,251.54)(0.492,-0.927){21}{\rule{0.119pt}{0.833pt}}
\multiput(243.17,253.27)(12.000,-20.270){2}{\rule{0.400pt}{0.417pt}}
\multiput(256.58,230.54)(0.493,-0.616){23}{\rule{0.119pt}{0.592pt}}
\multiput(255.17,231.77)(13.000,-14.771){2}{\rule{0.400pt}{0.296pt}}
\multiput(269.00,215.92)(0.539,-0.492){21}{\rule{0.533pt}{0.119pt}}
\multiput(269.00,216.17)(11.893,-12.000){2}{\rule{0.267pt}{0.400pt}}
\multiput(282.00,203.93)(0.669,-0.489){15}{\rule{0.633pt}{0.118pt}}
\multiput(282.00,204.17)(10.685,-9.000){2}{\rule{0.317pt}{0.400pt}}
\multiput(294.00,194.93)(0.950,-0.485){11}{\rule{0.843pt}{0.117pt}}
\multiput(294.00,195.17)(11.251,-7.000){2}{\rule{0.421pt}{0.400pt}}
\multiput(307.00,187.93)(1.033,-0.482){9}{\rule{0.900pt}{0.116pt}}
\multiput(307.00,188.17)(10.132,-6.000){2}{\rule{0.450pt}{0.400pt}}
\multiput(319.00,181.93)(1.378,-0.477){7}{\rule{1.140pt}{0.115pt}}
\multiput(319.00,182.17)(10.634,-5.000){2}{\rule{0.570pt}{0.400pt}}
\multiput(332.00,176.93)(1.378,-0.477){7}{\rule{1.140pt}{0.115pt}}
\multiput(332.00,177.17)(10.634,-5.000){2}{\rule{0.570pt}{0.400pt}}
\multiput(345.00,171.95)(2.472,-0.447){3}{\rule{1.700pt}{0.108pt}}
\multiput(345.00,172.17)(8.472,-3.000){2}{\rule{0.850pt}{0.400pt}}
\multiput(357.00,168.95)(2.695,-0.447){3}{\rule{1.833pt}{0.108pt}}
\multiput(357.00,169.17)(9.195,-3.000){2}{\rule{0.917pt}{0.400pt}}
\multiput(370.00,165.95)(2.472,-0.447){3}{\rule{1.700pt}{0.108pt}}
\multiput(370.00,166.17)(8.472,-3.000){2}{\rule{0.850pt}{0.400pt}}
\multiput(382.00,162.95)(2.695,-0.447){3}{\rule{1.833pt}{0.108pt}}
\multiput(382.00,163.17)(9.195,-3.000){2}{\rule{0.917pt}{0.400pt}}
\put(395,159.17){\rule{2.500pt}{0.400pt}}
\multiput(395.00,160.17)(6.811,-2.000){2}{\rule{1.250pt}{0.400pt}}
\put(407,157.17){\rule{2.700pt}{0.400pt}}
\multiput(407.00,158.17)(7.396,-2.000){2}{\rule{1.350pt}{0.400pt}}
\put(420,155.67){\rule{3.132pt}{0.400pt}}
\multiput(420.00,156.17)(6.500,-1.000){2}{\rule{1.566pt}{0.400pt}}
\put(433,154.17){\rule{2.500pt}{0.400pt}}
\multiput(433.00,155.17)(6.811,-2.000){2}{\rule{1.250pt}{0.400pt}}
\put(445,152.67){\rule{3.132pt}{0.400pt}}
\multiput(445.00,153.17)(6.500,-1.000){2}{\rule{1.566pt}{0.400pt}}
\put(458,151.17){\rule{2.500pt}{0.400pt}}
\multiput(458.00,152.17)(6.811,-2.000){2}{\rule{1.250pt}{0.400pt}}
\put(470,149.67){\rule{3.132pt}{0.400pt}}
\multiput(470.00,150.17)(6.500,-1.000){2}{\rule{1.566pt}{0.400pt}}
\put(483,148.67){\rule{3.132pt}{0.400pt}}
\multiput(483.00,149.17)(6.500,-1.000){2}{\rule{1.566pt}{0.400pt}}
\put(496,147.67){\rule{2.891pt}{0.400pt}}
\multiput(496.00,148.17)(6.000,-1.000){2}{\rule{1.445pt}{0.400pt}}
\put(508,146.67){\rule{3.132pt}{0.400pt}}
\multiput(508.00,147.17)(6.500,-1.000){2}{\rule{1.566pt}{0.400pt}}
\put(521,145.67){\rule{2.891pt}{0.400pt}}
\multiput(521.00,146.17)(6.000,-1.000){2}{\rule{1.445pt}{0.400pt}}
\put(533,144.67){\rule{3.132pt}{0.400pt}}
\multiput(533.00,145.17)(6.500,-1.000){2}{\rule{1.566pt}{0.400pt}}
\put(546,143.67){\rule{2.891pt}{0.400pt}}
\multiput(546.00,144.17)(6.000,-1.000){2}{\rule{1.445pt}{0.400pt}}
\put(571,142.67){\rule{3.132pt}{0.400pt}}
\multiput(571.00,143.17)(6.500,-1.000){2}{\rule{1.566pt}{0.400pt}}
\put(584,141.67){\rule{2.891pt}{0.400pt}}
\multiput(584.00,142.17)(6.000,-1.000){2}{\rule{1.445pt}{0.400pt}}
\put(558.0,144.0){\rule[-0.200pt]{3.132pt}{0.400pt}}
\put(609,140.67){\rule{2.891pt}{0.400pt}}
\multiput(609.00,141.17)(6.000,-1.000){2}{\rule{1.445pt}{0.400pt}}
\put(596.0,142.0){\rule[-0.200pt]{3.132pt}{0.400pt}}
\put(634,139.67){\rule{2.891pt}{0.400pt}}
\multiput(634.00,140.17)(6.000,-1.000){2}{\rule{1.445pt}{0.400pt}}
\put(621.0,141.0){\rule[-0.200pt]{3.132pt}{0.400pt}}
\put(659,138.67){\rule{3.132pt}{0.400pt}}
\multiput(659.00,139.17)(6.500,-1.000){2}{\rule{1.566pt}{0.400pt}}
\put(646.0,140.0){\rule[-0.200pt]{3.132pt}{0.400pt}}
\put(684,137.67){\rule{3.132pt}{0.400pt}}
\multiput(684.00,138.17)(6.500,-1.000){2}{\rule{1.566pt}{0.400pt}}
\put(672.0,139.0){\rule[-0.200pt]{2.891pt}{0.400pt}}
\put(722,136.67){\rule{3.132pt}{0.400pt}}
\multiput(722.00,137.17)(6.500,-1.000){2}{\rule{1.566pt}{0.400pt}}
\put(697.0,138.0){\rule[-0.200pt]{6.022pt}{0.400pt}}
\put(747,135.67){\rule{3.132pt}{0.400pt}}
\multiput(747.00,136.17)(6.500,-1.000){2}{\rule{1.566pt}{0.400pt}}
\put(735.0,137.0){\rule[-0.200pt]{2.891pt}{0.400pt}}
\put(797,134.67){\rule{3.132pt}{0.400pt}}
\multiput(797.00,135.17)(6.500,-1.000){2}{\rule{1.566pt}{0.400pt}}
\put(760.0,136.0){\rule[-0.200pt]{8.913pt}{0.400pt}}
\put(835,133.67){\rule{3.132pt}{0.400pt}}
\multiput(835.00,134.17)(6.500,-1.000){2}{\rule{1.566pt}{0.400pt}}
\put(810.0,135.0){\rule[-0.200pt]{6.022pt}{0.400pt}}
\put(885,132.67){\rule{3.132pt}{0.400pt}}
\multiput(885.00,133.17)(6.500,-1.000){2}{\rule{1.566pt}{0.400pt}}
\put(848.0,134.0){\rule[-0.200pt]{8.913pt}{0.400pt}}
\put(948,131.67){\rule{3.132pt}{0.400pt}}
\multiput(948.00,132.17)(6.500,-1.000){2}{\rule{1.566pt}{0.400pt}}
\put(898.0,133.0){\rule[-0.200pt]{12.045pt}{0.400pt}}
\put(1011,130.67){\rule{3.132pt}{0.400pt}}
\multiput(1011.00,131.17)(6.500,-1.000){2}{\rule{1.566pt}{0.400pt}}
\put(961.0,132.0){\rule[-0.200pt]{12.045pt}{0.400pt}}
\put(1074,129.67){\rule{3.132pt}{0.400pt}}
\multiput(1074.00,130.17)(6.500,-1.000){2}{\rule{1.566pt}{0.400pt}}
\put(1024.0,131.0){\rule[-0.200pt]{12.045pt}{0.400pt}}
\put(1162,128.67){\rule{3.132pt}{0.400pt}}
\multiput(1162.00,129.17)(6.500,-1.000){2}{\rule{1.566pt}{0.400pt}}
\put(1087.0,130.0){\rule[-0.200pt]{18.067pt}{0.400pt}}
\put(1250,127.67){\rule{3.132pt}{0.400pt}}
\multiput(1250.00,128.17)(6.500,-1.000){2}{\rule{1.566pt}{0.400pt}}
\put(1175.0,129.0){\rule[-0.200pt]{18.067pt}{0.400pt}}
\put(1351,126.67){\rule{3.132pt}{0.400pt}}
\multiput(1351.00,127.17)(6.500,-1.000){2}{\rule{1.566pt}{0.400pt}}
\put(1263.0,128.0){\rule[-0.200pt]{21.199pt}{0.400pt}}
\put(1364.0,127.0){\rule[-0.200pt]{18.067pt}{0.400pt}}
\put(181.0,123.0){\rule[-0.200pt]{0.400pt}{177.543pt}}
\put(181.0,123.0){\rule[-0.200pt]{303.052pt}{0.400pt}}
\put(1439.0,123.0){\rule[-0.200pt]{0.400pt}{177.543pt}}
\put(181.0,860.0){\rule[-0.200pt]{303.052pt}{0.400pt}}
\end{picture}
\end{center}
\caption{Mixing on the conifold varying with distance of the hidden
brane, $y_1=\log(r_1/r_s)$, up to the mouth of the throat at
$y_1=16$. ($M$ is the number of fractional D5 branes wrapped on a
compact $S^3 \subset T^{1,1}$ at the tip of the throat.)}
\label{conifoldmixing}\end{figure}
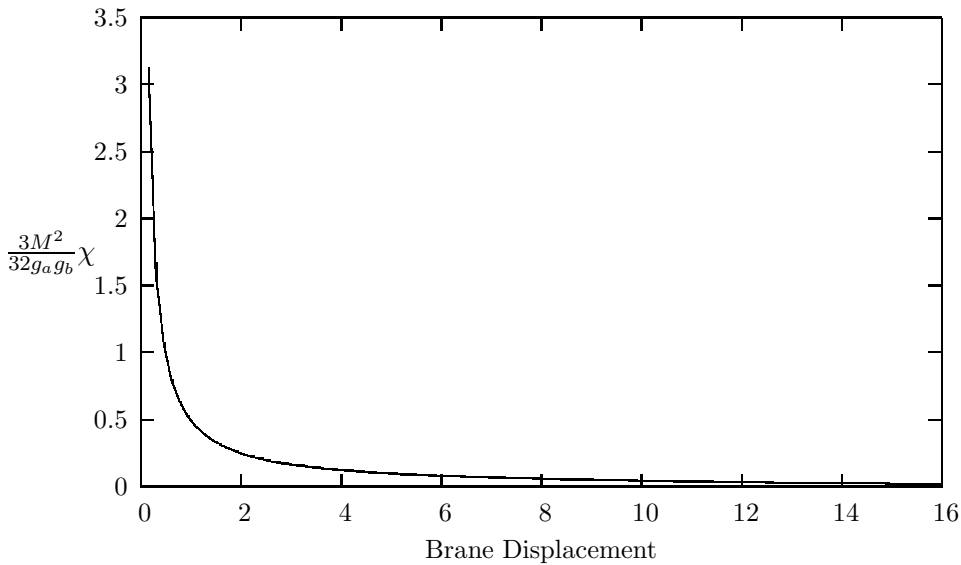

Using the analysis for the solution of Green functions, we find no contact terms on the brane,
and the boundary condition is just that $\partial_y \phi = 0$ at the branes. We then find
\beq
\chi_{ab} = g_{a}g_{b}\frac{32}{3 M^2} \frac{1}{4y_1 + 2(y_1 -y_0)^2)}
\frac{\tilde{G}_{<} (y_s)}{\tilde{G}_{<}^{\prime} (y_1)},
\eeq
where the hidden brane is placed at $y_1=\log(r_1/r_s)$.
The homogeneous solutions can be easily found numerically; a graph is given in
Fig.~\ref{conifoldmixing}. Again, the mixing is much larger than might naively have been expected;
the Klebanov-Tseytlin
throat is comparable to the RS model in the $mR\ll 1$ limit. Once the backreaction of the fluxes
becomes important, one might expect an exponential damping of the propagation similar to the $mR\gg 1$
limit of the RS model. Therefore in general warped throat configurations
it is not possible to place an upper or lower bound on the size of the kinetic mixing between branes:
there may indeed be an experimental signal to observe.

\section{Conclusions}

We have shown that models with
massless hidden $U(1)$s can be found in string theory, and argued that
they are natural for
certain classes of backgrounds. These
massless hidden-sector $U(1)$s can nevertheless have observable
and experimentally testable effects because they will typically
mix with the ordinary photon via a so-called kinetic mixing term. Using conformal field theory
and supergravity techniques we have calculated these effects.
The latter method can be used even when fluxes are included to stabilise the moduli. Moreover,
we have demonstrated that in general kinetic mixing is non-zero even
if all the $U(1)$s involved are anomaly free and therefore massless. This facilitates extremely
sensitive tests in a variety of current and near future
low-energy experiments.

The size of the kinetic mixing is model-dependent. Yet, for generic parameter values, it is often within
reach of current and near future experiments.
There is thus the real possibility
that an experimental signal will be observed soon that would give deep insights into the particular
string theory background upon
which we may live.
Alternatively, new stronger bounds will crucially exclude many models.
We believe that it is thus worthwhile to examine future models for such fields and
the kinetic mixing between them.

\bigskip
\bigskip

\centerline{\bf \large Acknowledgements}
\bigskip

MDG would like to sincerely thank the IPPP, Durham for hospitality.
AR would like to thank
Ralph Blumenhagen, Luis Ibanez, Jan Louis, Dieter L\"ust, Fernando Quevedo and Stefan Theisen
for enlightening discussions on hidden $U(1)$s in type II string phenomenology.

\bigskip

\appendix

\section{Remarks on the D3-\dbar{3} system}
\label{D3D3BRemarks}

We may wish to follow the procedure outlined in the section \ref{sec:CFT} for constructing a
massless $U(1)$ field from two stacks of branes applied to the D3-$\dthreebar$ system, and start
with two stacks: one of two D3-branes, one of two \dbar{3}-branes. We consider the compact space
to be a six-torus, but the discussion regarding the masses applies to \emph{any} manifold. We then
split these into four stacks, giving four gauge fields
$A^{\alpha} = \{A_a^1,A_{a}^2,A_{\bar{b}}^1,A_{\bar{b}}^2\}$. Due to the mutual supersymmetries preserved,
there is only kinetic/mass mixing between the branes and antibranes, not amongst themselves. However,
a crucial difference between this system and one of purely branes is that there are uncancelled NS-NS
tadpoles, and thus we have a non-zero contribution to the mass from the planar diagrams (with both vertex
operators on one boundary). The planar and non-panar  masses are given by
\begin{eqnarray}
m_{\rm planar}^2 &=& \tr(\lambda_a^i \lambda_{a}^i) \sum_j \tr(\gamma_{\bar{b}}^j) m^2, \nonumber \\
m_{\rm non-planar}^2 &=& \tr(\lambda_a^i ) \tr(\lambda_{\bar{b}}^j) m^2,
\end{eqnarray}
where $m^2$ is given by equation (\ref{m2}). Thus we can write the Lagrangian as
\beq
\mathcal{L} \supset \frac{1}{2} (A_{\mu})^{\alpha} (\mathcal{M} )_{\alpha \beta} (A^{\mu})^{\beta}
- \frac{1}{4 g^2} (F_{\mu \nu})^{\alpha} (\mathcal{X} )_{\alpha \beta} (F^{\mu \nu})^{\beta},
\eeq
where
\beq
\mathcal{M} = \left( \begin{array}{cccc} 4m^2 & 0 & m^2 & m^2 \\ 0 &4m^2 & m^2 & m^2
\\ m^2 & m^2 & 4m^2 & 0 \\ m^2 & m^2 & 0 & 4 m^2 \end{array} \right)
\eeq
and
\beq
\mathcal{X} = \left( \begin{array}{cccc} 1 & 0 & -\chi_{11} & -\chi_{12} \\ 0 & 1 & -\chi_{21} &
-\chi_{22} \\  -\chi_{11} & -\chi_{21} & 1 & 0 \\ -\chi_{12} & -\chi_{22} & 0 & 1 \end{array} \right).
\eeq
Note that upon diagonalising $\mathcal{M}$, we find \emph{no} massless $U(1)$s; all four fields become
massive, with masses multiples of $m^2$. Note also the presence of the diagonal mass terms from the
planar diagrams. This is a new feature present when we have broken supersymmetry, and if we analyse the
supergravity calculation we find that it arises from a dilaton tadpole. This occurs because of the
uncancelled NS-NS charges present.


\end{document}